\documentclass[12pt]{article}

\usepackage[margin=2.5cm]{geometry}
\usepackage{amsthm,amsmath,amsfonts,amssymb}
\usepackage{graphicx,float}
\usepackage{multirow,setspace}
\usepackage[authoryear]{natbib}
\usepackage{enumerate}
\usepackage{caption}
\usepackage{subcaption}
\usepackage{bm}
\usepackage{color}
\usepackage{xcolor}%
\usepackage{url}
\usepackage[colorlinks,citecolor=blue,urlcolor=blue]{hyperref}
\usepackage[format=plain, labelfont={bf,it}, textfont=it]{caption}
\usepackage{listings}%
\usepackage{lstbayes}
\usepackage{booktabs}
\theoremstyle{plain}

\newtheorem{theorem}{Theorem}[section]

\theoremstyle{remark}

\title{\vspace{-40pt}Objective Priors for the Conway-Maxwell-Poisson (COM-Poisson) Distribution\vspace{-10pt}}
\author{Mark J. Meyer \\ Department of Mathematics and Statistics, Georgetown University \\ Washington, DC USA\\
Amia Graye \\ Department of Mathematics and Statistics, Georgetown University \\ Washington, DC USA \\
Kimberly F. Sellers \\ Department of Statistics, North Carolina State University \\ Raleigh, NC USA}
\date{}

\begin{document}

	\maketitle
	
\begin{abstract}
The Conway-Maxwell-Poisson (COM-Poisson) distribution is a flexible, two parameter distribution for count data that can accommodate equidispersion, overdispersion, and underdispersion in the data. Previous Bayesian evaluations of the COM-Poisson distribution have largely focused on the regression context with little discussion on objective priors for the base model. Additionally, we find the multivariate Jeffreys' prior is inadequate for this model. Motivated by this, we propose four different objective priors when using the COM-Poisson distribution including an independent Jefferys' prior model and a novel reference prior derived using the hierarchical approach for objective priors. Since the resulting reference prior is non-standard, we implement constrained nonlinear optimization by linear approximation to obtain the reference prior parameters for an Empirical Bayes approach. We also develop \texttt{Stan} code to perform No U-Turn Sampling on each model. We compare all four approaches in simulation and demonstrate the models in several data illustrations that span the spectrum of dispersion types.
\end{abstract}

\section{Introduction}
\label{s:intro}

The Conway-Maxwell-Poisson (COM-Poisson) distribution is a two-parameter flexible model derived that can describe the variation of count data \citep{Conway1962}. While several parametrizations of the COM-Poisson distribution have been proposed \citep[Chapter 2]{Sellers2023}, this work assumes the COM-Poisson parametrization. This distribution has a probability mass function of the form
\begin{eqnarray} \label{cmppmf}
P(X=x) = \frac{\lambda^x}{(x!)^\nu Z(\lambda, \nu)}, \hspace{0.5in} x = 0, 1, 2, \ldots,
\end{eqnarray}
for a random variable $X$, where $\lambda = E(X^\nu) >0$ is a generalized form of the Poisson rate parameter, $\nu \ge 0$ is a dispersion parameter, and $Z(\lambda, \nu) = \sum_{j=0}^{\infty} \frac{\lambda^j}{(j!)^\nu}$ is the normalizing function. The dispersion parameter $\nu \ge 0$ is such that $\nu = 1$ denotes equidispersion, and $\nu > (<) 1$ signifies underdispersion (overdispersion) relative to the Poisson model. Given a sample of observations $x_1, \ldots, x_n$ from a COM-Poisson($\lambda, \nu$) distribution, the likelihood and log-likelihood functions are 
\begin{align} 
{\mathcal L}(\lambda, \nu | x_1, \ldots, x_n )&= \prod_{i=1}^{n} \frac{\lambda^{x_i}}{(x_i!)^\nu Z(\lambda, \nu)}
= \frac{\lambda^{\sum_{i=1}^{n} x_i}}{[Z(\lambda, \nu)]^n \left(\prod_{i=1}^{n} x_i! \right)^\nu}\label{eq:like} \\  \text{ and} \nonumber\\ 
\ell(\lambda, \nu | x_1, \ldots, x_n ) &= \sum_{i=1}^{n} x_i \ln \lambda - \nu \sum_{i=1}^{n} \ln (x_i!) - n\ln Z(\lambda, \nu)\label{eq:loglike}
\end{align}
where ${\mathcal L}(\cdot)$ and $\ell(\cdot)$ respectively denote the likelihood and log-likelihood. The COM-Poisson distribution is a flexible model that contains three well-known distributions. 
Two of the special case distributions are the Poisson distribution with rate parameter $\lambda$ (when $\nu=1$) and the geometric distribution with success probability $1-\lambda$ (when $\nu=0$ and $\lambda < 1$). 
The Bernoulli distribution with success probability $p=\frac{\lambda}{1+\lambda}$ is meanwhile a limiting case for $\nu \rightarrow \infty$. Table~\ref{t:special} provides details regarding these three cases, given their corresponding forms for $Z(\lambda, \nu)$ under each constraint. 

\begin{table} 
\centering
\caption{Special cases corresponding to the COM-Poisson($\lambda, \nu$) distribution. For each constraint regarding $\nu$, this table presents the corresponding $Z(\lambda, \nu)$, distribution, and corresponding Jeffreys' prior under the COM-Poisson parameterization.} \label{t:special}
\begin{tabular}{| l | l | l | l |} 
\hline
Special cases & $Z(\lambda, \nu)$ & Distribution & Jeffreys' Prior\\
\hline
$\nu=1$ & $e^\lambda$ & Poisson($\lambda)$ & $\propto \lambda^{-1}$\\
$\nu=0$, $0 < \lambda < 1$ & $\frac{1}{1-\lambda}$ & Geometric($p=1-\lambda$) & $\propto(1-\lambda)^{-1} \lambda^{-1/2}$\\
$\nu \rightarrow \infty$ & $1+\lambda$ & Bernoulli$\left(p=\frac{\lambda}{1+\lambda} \right)$ & $\propto\left(\frac{\lambda}{1+\lambda} \right)^{1/2 - 1} \left(\frac{1}{1+\lambda} \right)^{1/2 - 1}$\\ \hline
\end{tabular}
\end{table}

Bayesian evaluations of (or stemming from) this model include works on the conjugate prior for the base model \citep{Kadane2006} and numerous regression-related methods \citep{Chanialidis2018,Huang2021,Benson2021}. As an exponential family, a conjugate prior for the COM-Poisson exists, subject to a set of constraints on the prior parameters \citep{Kadane2006}. This work demonstrates the relationship between the conjugate prior and the special cases, showing that the expected conjugate priors arise in each case. The resulting distribution is non-standard and while some discussion of parameter elicitation is included, the authors do not discuss objective prior specifications. Others consider an ``intercept-only'' regression model in a data illustration, using the mean-parameterization of the COM-Poisson with a Gamma$(1,1)$ distribution on the mean parameter and a Gamma$(0.0625, 0.25)$ on the dispersion parameter \citep{Benson2021}. The prior mean on the dispersion parameter is $1/4$, meaning this prior imparts some information, suggesting over-dispersion in the data. The work does not, however, consider a joint prior on $\nu$ and $\lambda$. In the regression setting more generally, normal priors with large variances have been selected for the components of the mean model, e.g. a mean zero normal with variance equal to $10^6$ \citep{Chanialidis2018}, $10^5$ \citep{Huang2021}, or $5^2$ \citep{Benson2021}. A log-normal prior with mean equal to zero and variance set to $10^5$ for $\nu$ has also been used \citep{Huang2021}. These priors were all considered in data illustrations, and their properties were not explored. The regression-based work on the COM-Poisson model implements commonly used diffuse priors on the model components. However, to our knowledge, no previous work examines objective priors for the base COM-Poisson model whose likelihood is described in Equation~\eqref{eq:like}.

Objective priors are typically derived from the assumed model as opposed to a prior that is developed based on historical data or so-called ``expert'' knowledge. They have a rich history of study given the need for prior selection in Bayesian models when it is difficult to accurately elicit a prior distribution. One of the most common choices of objective prior is the reference prior \citep{Bernardo1979, Berger1989, BergerBernardo1992BS4, BergerBernardo1992, Sun1998, Bernardo2005, Berger2009, Ghosh2011, Berger2012, Berger2015}. In the univariate case, the reference prior often reduces to the well known Jeffreys' prior which is based on the information \citep{Jeffreys1946}. The COM-Poisson model is generally a two parameter model but it reduces to a single parameter under the three special cases: $\nu = 1$ (Poisson), $\nu = 0$ (Geometric), and $\nu \rightarrow \infty$ (Bernoulli). As Table\ref{t:special} illustrates, the univariate Jeffreys' prior derived from the special case formulation of the likelihood in Equation~\eqref{eq:like} is equivalent to the corresponding Jeffreys' prior for the special case distribution itself. Thus, if $\nu$ was known or highly suspected to be near one of the special cases, one could simple use the univariate Jeffreys' prior as the objective prior. When $\nu$ is unknown, we must turn to a multivariate objective prior.

In the multivariate settings, Jeffreys' prior does not guarantee objectivity \citep{BergerBernardo1992,Ghosh2011}. It may also be difficult to work with. In our case, the information matrix for the COM-Poisson model is
\begin{align*}
	\mathcal{I}(\lambda, \nu) = \left[ \begin{array}{cc}
		\frac{n}{\lambda} \frac{\partial}{\partial \lambda} \ln\left\{ Z(\lambda, \nu) \right\} + n \frac{\partial^2}{\partial \lambda^2} \ln\left\{ Z(\lambda, \nu) \right\}  & n \frac{\partial^2}{\partial\nu\partial \lambda} \ln\left\{ Z(\lambda, \nu) \right\} \\ 
		n \frac{\partial^2}{\partial\nu\partial \lambda} \ln\left\{ Z(\lambda, \nu) \right\} & n\frac{\partial^2}{\partial\nu^2} \ln\left\{ Z(\lambda, \nu) \right\} \\
	\end{array} \right],
\end{align*}
with a resulting multivariate Jeffreys' prior of 
	\begin{align}
		J(&\lambda, \nu) \propto \nonumber \\
		&\sqrt{\left[ \frac{1}{\lambda} \frac{\partial}{\partial \lambda} \ln\left\{ Z(\lambda, \nu) \right\}+  \frac{\partial^2}{\partial \lambda^2 }  \ln\left\{ Z(\lambda, \nu) \right\}  \right] \frac{\partial^2}{\partial\nu^2} \ln\left\{ Z(\lambda, \nu) \right\} - \left[ \frac{\partial^2}{\partial\nu\partial \lambda} \ln\left\{ Z(\lambda, \nu) \right\} \right]^2}. \label{eq:jeffy}
	\end{align}
	There are several issues posed by this prior. First, we must work with the first and second derivatives of the normalizing constant which contains an infinite sum. Second, the expression under the root is not guaranteed to be positive. And third, its formulation indicates some level of prior dependence between $\nu$ and $\lambda$ that we may not wish to impose. This prior model's deficiencies thusly serves as motivation for the work at hand where we develop and compare four different objective priors for the COM-Poisson base model. Our priors of interest are a joint flat prior, a joint limiting prior, an independent Jeffreys' prior, and a hierarchical reference prior.

\section{Statistical Framework}
\label{s:statframe}

The likelihood, as derived in Section~\ref{s:intro}, has the form
\begin{align*} \nonumber
	{\mathcal L}(\lambda, \nu; {\bm{x}}) = \lambda^{\sum_{i=1}^n x_i} e^{-\nu \sum_{i=1}^n \log(x_i!)} \left[Z(\lambda, \nu)\right]^{-n}.
\end{align*}
Noting that the COM-Poisson base model is an exponential family member, prior work on the model has established a conjugate prior \citep{Kadane2006}. For positive and real-valued hyper-parameters $a, b,$ and $c$, the form of that prior is
\begin{align}
	\pi(\lambda, \nu) = \kappa(a, b, c) \lambda^{a - 1}e^{-\nu b} \left[Z(\lambda, \nu)\right]^{-c}, \label{eq:cp}
\end{align}
where the normalizing constant, $\kappa(a, b, c)$, is given by
\begin{align}
	\kappa(a, b, c)^{-1} = \int_0^{\infty}\int_0^{\infty} \lambda^{a - 1}e^{-\nu b} \left[Z(\lambda, \nu)\right]^{-c} d\lambda d\nu. \label{eq:kappa}
\end{align}
The double integral in Equation~\eqref{eq:kappa} converges for $a, b, c > 0$ subject to the constraint
\begin{align}
		\frac{b}{c} > \ln\left(\left\lfloor \frac{a}{c} \right\rfloor !\right) + \left( \frac{a}{c} - \left\lfloor \frac{a}{c} \right\rfloor \right)\ln\left(\left\lfloor \frac{a}{c} \right\rfloor + 1 \right), \label{eq:con}
	\end{align}
where $\lfloor\cdot\rfloor$ denotes the floor function \citep{Kadane2006}. Using the prior in Equation~\eqref{eq:cp} results in a general posterior of the form
	\begin{align}
		p(\lambda, \nu | x_1, \ldots, x_n) \propto \lambda^{a + \sum_{i=1}^n x_i - 1} e^{-\nu \left[b + \sum_{i=1}^n \ln(x_i!) \right]} \left[Z(\lambda, \nu)\right]^{-(c+n)}. \label{eq:conjPost}
	\end{align}
	
	The upper bound on Equation~\eqref{eq:kappa} is of particular interest to us. We can decompose the normalizing constant into two components \citep{Kadane2006}:
\begin{align}
	\kappa(a, b, c)^{-1} = I_1 + I_2, \label{eq:kappaDecomp}
\end{align}
where $I_1$ and $I_2$ are integrals of the form
\begin{align*}
	I_1 &= \int_{0}^{\infty} e^{-b\nu} \int_{0}^1 \lambda^{a - 1} \left[Z(\lambda, \nu)\right]^{-c} d\lambda d\nu\\ 
	\text{ and}\\
	I_2 &= \int_{0}^{\infty} e^{-b\nu} \int_{1}^{\infty} \lambda^{a - 1} \left[Z(\lambda, \nu)\right]^{-c} d\lambda d\nu.
\end{align*}
Both $I_1$ and $I_2$ are bounded above. Beginning with $I_1$,
\begin{align*}
	I_1 &\leq \int_{0}^{\infty} e^{-b\nu} \int_{0}^1 \lambda^{a - 1} d\lambda d\nu.
\end{align*}
To establish the upper bound on $I_2$, define $Q$ to be a random variable on the non-negative integers with mass function $P(Q = j) = q_j$ and thus $\sum_{j=0}^{\infty} q_j = 1$. Via an application of Jensen's inequality, we can obtain a lower bound on $Z(\lambda, \nu)$ of the form
\begin{align*}
	Z(\lambda, \nu) \geq \lambda^{E(Q)} e^{-E[\log(Q!)]\nu} \prod_{j=0}^{\infty} q_j^{q_j}.
\end{align*}
The upper bound on $I_2$ is then
\begin{align*}
	I_2 \leq \int_{0}^{\infty} e^{-\nu \left[ b - c E\{\log(Q!)\} \right]} dv \int_{1}^{\infty} \lambda^{-[c E(Q) - a + 1]}d\lambda  \prod_{j=0}^{\infty} q_j^{-c q_j},
\end{align*}
where for some $\epsilon > 0$, 
\begin{align}
	E(Q) & =  \frac{a}{c} + \epsilon, \label{eq:EQ}\\
	E\{ \log(Q!) \} &= \log\left(\left\lfloor \frac{a}{c} \right\rfloor !\right) + \left( \frac{a}{c} - \left\lfloor \frac{a}{c}  \right\rfloor + \epsilon \right)\ln\left(\left\lfloor \frac{a}{c} \right\rfloor + 1 \right),\label{eq:ElogQ}
\end{align}
and
\begin{align}
	q_j = \left\{ \begin{array}{cl}
			1 - \left( \frac{a}{c} - \left\lfloor \frac{a}{c} \right\rfloor + \epsilon \right) & \text{if } j = \left\lfloor \frac{a}{c} \right\rfloor \\
			\frac{a}{c} - \left\lfloor \frac{a}{c} \right\rfloor + \epsilon & \text{if } j = \left\lfloor \frac{a}{c} \right\rfloor + 1\\
			0 & \text{o/w}
		\end{array}\right. . \label{eq:qj}
\end{align}
For more information on the derivation of the bounds on $I_1$ and $I_2$, see the seminal work on the conjugate prior for the base COM-Poisson model \citep{Kadane2006}.

	We now explicitly derive the upper bounds of both $I_1$ and $I_2$ to obtain a closed form bound on $\kappa(a,b,c)^{-1}$. Beginning with $I_1$, and evaluating the inner integral, we have that
\begin{align}
	I_1 &\leq  \int_{0}^{\infty} e^{-b\nu} \left[ \frac{\lambda^a}{a} \right]_0^1 d\nu = \frac{1}{a} \int_{0}^{\infty} e^{-b\nu} d\nu = \frac{1}{ab}, \label{eq:I1}
\end{align}
where the remaining integral is the kernel of an exponential density integrated over its support. The upper bound on $I_1$ holds for $a, b > 0$. Working from $I_2$, we see that
\begin{align}
	I_2 &\leq \int_{0}^{\infty} e^{-\nu \left[ b - c E\{\log(Q!)\} \right]} dv \int_{1}^{\infty} \lambda^{-[c E(Q) - a + 1]}d\lambda  \prod_{j=0}^{\infty} q_j^{-c q_j} \nonumber\\
	&= \frac{1}{\left[b - c E\{\log(Q!)\} \right] } \frac{1}{\left[c E(Q) - a\right]} \prod_{j=0}^{\infty} q_j^{-c q_j}. \label{eq:I2}
\end{align}
The first integral is the kernel of an exponential density function and holds for $b > c E\{\log(Q!)\} $. The second integral is the kernel of a Pareto density function with minimum equal to 1 and shape $c E(Q) - a$. Integration over its support holds for $c E(Q) > a$. The forms for $E(Q)$, $E\{\log(Q!)\}$, and the $q_j$ are as described in Equations~\eqref{eq:EQ} to~\eqref{eq:qj}. Combining Equations~\eqref{eq:kappaDecomp} with~\eqref{eq:I1} and~\eqref{eq:I2}, we obtain an explicit upper bound on $\kappa(a,b,c)^{-1}$:
\begin{align}
	\kappa(a,b,c)^{-1} \leq \frac{1}{ab} + \frac{1}{\left[b - c E\{\log(Q!)\} \right] } \frac{1}{\left[c E(Q) - a\right]} \prod_{j=0}^{\infty} q_j^{-c q_j}. \label{eq:kappaBound}
\end{align}
This bound holds provided $a,b,$ and $c$ satisfy the constraint in Equation~\eqref{eq:con}.

\section{Objective Priors}
\label{s:priors}

We now describe four different choices of objective priors for the COM-Poisson base model. The joint flat prior, joint limiting prior, and hierarchical reference prior will all lead to a proper posterior, as we will demonstrate. While we cannot guarantee a proper posterior when using the independent Jeffreys' prior, we can numerically evaluate it.

\subsection{Joint Flat Prior}

Assuming a complete absence of information, the na\"ive choice of joint prior on $\lambda$ and $\nu$ is the joint flat prior:
\begin{align}
	\pi(\lambda, \nu) \propto 1, \label{eq:flatPrior}
\end{align}
which is an improper prior. Using the prior in Equation~\eqref{eq:flatPrior}, the resulting posterior is
\begin{align}
	\pi(\lambda, \nu | x_1, \ldots, x_n ) \propto \lambda^{\sum_{i=1}^n x_i + 1 - 1} e^{-\nu \sum_{i=1}^n \log(x_i!)} \left[Z(\lambda, \nu)\right]^{-n}. \label{eq:flatPost}
\end{align}

\begin{theorem}\label{th:flat}
	If the observed data, $\bm{x}$, satisfies the condition that
	\begin{align*}
		\frac{\sum_{i=1}^n \log(x_i!)}{n} > \ln\left(\left\lfloor \bar{x} + \frac{1}{n} \right\rfloor !\right) + \left( \bar{x} + \frac{1}{n} - \left\lfloor \bar{x} + \frac{1}{n} \right\rfloor \right)\ln\left(\left\lfloor \bar{x} + \frac{1}{n} \right\rfloor + 1 \right),
	\end{align*}
	then the posterior in Equation~\eqref{eq:flatPost} is proper.
\end{theorem}
The proof is immediate, provided the data satisfies the condition. This is because the resulting posterior has the same distribution as the conjugate prior which is proper under the condition in Equation~\eqref{eq:con}.

\subsection{Joint Limiting Prior}

Consider again the resulting posterior when using the conjugate prior:
\begin{align*}
	p(\lambda, \nu | x_1, \ldots, x_n) \propto \lambda^{a + \sum_{i=1}^n x_i - 1} e^{-\nu \left[b + \sum_{i=1}^n \ln(x_i!) \right]} \left[Z(\lambda, \nu)\right]^{-(c+n)}.
\end{align*}
Interpreting the the hyper-parameters of the conjugate prior in Equation~\eqref{eq:cp} as ``added information,'' we note that $a$ can be interpreted as an additional data point $x'$ or additional sum of data points $x_j'$ for $j = 1, \ldots, m$. The parameter $c$ has a similarly straightforward interpretation as the additional number of data points used to construct the prior, i.e. the value $m$. The final parameter, $b$, is harder to interpret but is the additional sum of the (natural) log of the factorial of the additional data points $x_j'$ for $j = 1, \ldots, m$. From these interpretations, if we let $a, b,$ and $c$ simultaneously tend toward zero we are limiting the additional information imparted by the conjugate prior. Obtaining the limit, the prior is
\begin{align}
	\lim_{(a, b, c) \rightarrow (0, 0, 0)} \lambda^{a - 1}e^{-\nu b} \left[Z(\lambda, \nu)\right]^{-c} = \lambda^{-1}, \label{eq:limitPrior}
\end{align}
which we refer to as a joint limiting prior. Using the prior in Equation~\eqref{eq:limitPrior}, the resulting posterior is
\begin{align}
	\pi(\lambda, \nu | x_1, \ldots, x_n ) \propto \lambda^{\sum_{i=1}^n x_i - 1} e^{-\nu \sum_{i=1}^n \log(x_i!)} \left[Z(\lambda, \nu)\right]^{-n}. \label{eq:limitPost}
\end{align}

\begin{theorem}\label{th:limit}
	If the observed data, $\bm{x}$, satisfies the condition that
	\begin{align*}
		\frac{\sum_{i=1}^n \log(x_i!)}{n} > \ln\left(\left\lfloor \bar{x} \right\rfloor !\right) + \left( \bar{x}- \left\lfloor \bar{x}\right\rfloor \right)\ln\left(\left\lfloor \bar{x} \right\rfloor + 1 \right),
	\end{align*}
	then the posterior in Equation~\eqref{eq:limitPost} is proper.
\end{theorem}
The proof is once again immediate, provided the data satisfies the stated condition. The joint limiting prior itself is equivalent to the Jeffreys' prior when $\nu = 1$ and the COM-Poisson model reduces to the Poisson case, see Table~\ref{t:special}. Using this prior is akin to assuming $\pi(\nu) \propto 1$ since it imparts no information regarding $\nu$. The difference between the resulting posteriors in Equations~\eqref{eq:flatPost} and~\eqref{eq:limitPost} is subtle and will be less noticeable when $\sum_{i=1}^n x_i$ is large.

\subsection{Independent Jeffreys' Prior}

As we discuss in Section~\ref{s:intro}, the joint multivariate Jeffreys' prior is problematic for a number of reasons \citep{Ghosh2011}. One alternative to working with Jeffreys' priors in a multivariate setting is to assume prior independence of the parameters and construct conditional reference priors based on the information for each parameter, separately. This approach is similar to a conditional prior approach based off of the information \citep{Sun1998}. Thus, rather than the prior having the form in Equation~\eqref{eq:jeffy}, it has the form
\begin{align}
	J(\lambda, \nu) \propto \sqrt{\left[ \frac{1}{\lambda} \frac{\partial}{\partial \lambda} \ln\left\{ Z(\lambda, \nu) \right\}+  \frac{\partial^2}{\partial \lambda^2 }  \ln\left\{ Z(\lambda, \nu) \right\}  \right] \frac{\partial^2}{\partial\nu^2} \ln\left\{ Z(\lambda, \nu) \right\} }. \label{eq:indyJeffy}
\end{align}
Since the COM-Poisson base model is a regular exponential family member this prior should at least be finite when the true parameters are on the interior of the parameter space. Using the prior in Equation~\eqref{eq:indyJeffy} results in the posterior 
\begin{align}
	\pi(\lambda, \nu | x_1, \ldots, x_n ) &\propto \lambda^{\sum_{i=1}^n x_i} e^{-\nu \sum_{i=1}^n \log(x_i!)} \left[Z(\lambda, \nu)\right]^{-n}\nonumber \\
	&\times \sqrt{\left[ \frac{1}{\lambda} \frac{\partial}{\partial \lambda} \ln\left\{ Z(\lambda, \nu) \right\}+  \frac{\partial^2}{\partial \lambda^2 }  \ln\left\{ Z(\lambda, \nu) \right\}  \right] \frac{\partial^2}{\partial\nu^2} \ln\left\{ Z(\lambda, \nu) \right\} }. \label{eq:indyPost}
\end{align}
We will evaluate this posterior numerically in our empirical experiments.

\subsection{Hierarchical Reference Prior}

A more formal objective prior can be obtained by employing the hierarchical approach to deriving the reference prior \citep{Berger2015}. This approaches shifts the problem of selecting a reference prior to a ``higher level'' of the model. We begin by selecting a class of proper priors, in this case the conjugate prior described in Equation~\eqref{eq:cp} where the parameters $a, b$, and $c$ are considered to be unknown. We then integrate out $\lambda$ and $\nu$ to obtain the marginal likelihood of the data depending only upon the hyper-parameters. The final step is to obtain estimates of $a, b$, and $c$ either computationally or, as computations in higher levels become more complex, using the Type-II maximum likelihood estimate when $n$ is large \citep{Berger2015}. The latter is results in an Empirical Bayes approach to selecting the prior parameters objectively.

Beginning with the posterior that results from using the conjugate prior, we integrate out $\lambda$ and $\nu$ to obtain the marginal likelihood:
\begin{align*}
	\int_0^{\infty} &\int_0^{\infty} \lambda^{\sum_{i=1}^n x_i} e^{-\nu \sum_{i=1}^n \ln(x_i!)} \left[Z(\lambda, \nu)\right]^{-n} \kappa(a, b, c) \lambda^{a - 1}e^{-\nu b} \left[Z(\lambda, \nu)\right]^{-c} d\lambda d\nu\\
		&= \kappa(a, b, c) \int_0^{\infty} \int_0^{\infty} \lambda^{\sum_{i=1}^n x_i + a - 1} e^{-\nu \left[b + \sum_{i=1}^n \ln(x_i!)\right]} \left[Z(\lambda, \nu)\right]^{-(n+c)}  d\lambda d\nu\\
		&= \frac{\kappa(a, b, c)}{\kappa\left[a + \sum_{i=1}^n x_i, b + \sum_{i=1}^n \ln(x_i!), c + n \right]}.
\end{align*}
Thus, the marginal likelihood is equivalent to the predictive distribution for the conjugate prior \citep{Kadane2006}. Inverting the $\kappa$ functions, we have
\begin{align*}
	p(x_1, \ldots, x_n | a, b, c) = \frac{\kappa^{-1}\left[a + \sum_{i=1}^n x_i, b + \sum_{i=1}^n \ln(x_i!), c + n \right]}{\kappa^{-1}(a, b, c)}.
\end{align*}
The distribution of the marginal likelihood is non-standard and difficult to sample from. Valid inference is also subject to the constraint on the parameters in Equation~\eqref{eq:con}. Since this is complex setting, we can select $a, b,$ and $c$ to be the values that maximize the marginal likelihood. However, direct maximization of the marginal likelihood is nontrivial. Thus, we establish an upper bound on it and maximize the bound. Let $B(a,b,c)$ be the function describing the bound on $\kappa^{-1}$ derived in Equation~\eqref{eq:kappaBound}. We can then bound the marginal likelihood:
\begin{align}
	p(x_1, \ldots, x_n | a, b, c) \leq \frac{B[a + \sum_{i=1}^n x_i, b + \sum_{i=1}^n \ln(x_i!), c + n]}{B(a,b,c)}.\label{eq:bound}
\end{align}
Maximizing Equation~\eqref{eq:bound} will require a constrained optimization. The resulting posterior when using this prior is the posterior in Equation~\eqref{eq:conjPost} that results from implementing the conjugate prior with $a, b$, and $c$ set to the values that maximize Equation~\eqref{eq:bound}. Let $\hat{a}, \hat{b}$, and $\hat{c}$ denote these values.
\begin{theorem}\label{th:limit}
	If the values $\hat{a}, \hat{b}$, and $\hat{c}$, satisfies the condition that
	\begin{align*}
		\frac{\hat{b}}{\hat{c}} > \ln\left(\left\lfloor \frac{\hat{a}}{\hat{c}} \right\rfloor !\right) + \left( \frac{\hat{a}}{\hat{c}} - \left\lfloor \frac{\hat{a}}{\hat{c}} \right\rfloor \right)\ln\left(\left\lfloor \frac{\hat{a}}{\hat{c}} \right\rfloor + 1 \right), \label{eq:con}
	\end{align*}
	then the hierarchical reference prior,
	\begin{align*}
		\pi(\lambda, \nu) = \kappa\left(\hat{a}, \hat{b}, \hat{c}\right) \lambda^{\hat{a} - 1}e^{-\nu \hat{b}} \left[Z(\lambda, \nu)\right]^{-\hat{c}},
	\end{align*}
	is proper. The resulting posterior model will also be proper.
\end{theorem}
\noindent The proof is again immediate provided the stated condition holds \citep{Kadane2006}.

\section{Implementation}
\label{s:code}

\subsection{Constrained Optimization}

Before generating posterior samples for the hierarchical reference prior model, we must first obtain Empirical Bayes estimates of $a, b$, and $c$ subject to the nonlinear constraint in Equation~\ref{eq:con}. In Section~\ref{s:priors}, we denote these values as $\hat{a}$, $\hat{b}$, and $\hat{c}$. To perform constrained optimization, we utilize the nonlinear optimization routine developed in the in the open-source \texttt{NLopt} library \citep{NLopt2008} as implemented by the \texttt{R} package \texttt{nloptr} \citep{nloptr2025}. Full code for eliciting prior parameters for the hierarchical reference prior is available online at \url{https://github.com/markjmeyer/CMP}. The implementation is quite fast even for large sample sizes and thus does not inhibit the use of the approach.

We must make several choices when implementing the optimization. First is the choice of some $\epsilon > 0$ which the upper bound depends on through Equations~\eqref{eq:EQ} to~\eqref{eq:qj}. For the bound to be finite, $\epsilon$ must be small enough \citep{Kadane2006}. Thus, we select $\epsilon = 1\times10^{-10}$. We must also select a tolerance for the optimization which we take to be small, also $1\times10^{-10}$. Next, we have to select initial values for $a, b$, and $c$. To obtain this, we randomly draw values on the interval $[0.2, 2]$ using a uniform distribution. These values must satisfy the constraint in Equation~\ref{eq:con}. If they do not, we repeat the draw until we obtain values that do. Finally, we select the algorithm for the optimizer. Since our optimization is subject to a nonlinear inequality constraint, we are limited in our choice and use Constrained Optimization BY Linear Approximations (COBYLA) for derivative-free optimization \citep{Powell1994}. This corresponds to the algorithm {\verb|NLOPT_LN_COBYLA|} in the \texttt{nloptr} \texttt{R} package \citep{nloptr2025}.

\subsection{Posterior Sampling}

The conjugate prior density, while recognizable as such, is a non-standard density, as is the resulting posterior when using  the independent Jeffreys' prior. Thus, regardless of prior specification, all resulting posterior densities are non-standard and require additional steps to obtain samples. The intractable nature of the likelihood makes sampling a challenge. The posterior of the COM-Poisson regression model is actually doubly-intractable due to the likelihood's normalizing constant in addition to the fact that the posterior cannot be normalized  \citep{Benson2021}. To handle this, previous authors consider a range of techniques. The original work on the conjugate prior treats $\nu$ as a nuisance parameter and describe steps to sample from the conditional distribution of $\lambda$ given $\nu$ under different scenarios, e.g. when $c > a$ in the conjugate prior or when $\nu < 1$ \citep{Kadane2006}. The sampler for the conditional distribution relies on a rejection algorithm with proposals for $\lambda$ drawn from either an $F$ distribution or a Gamma, depending on the setting. In the regression context, others employ an exchange algorithm \citep{Moller2006} with two steps aimed at reducing the correlation between successive samples of the coefficients from both the mean and dispersion model \citep{Chanialidis2018}. To improve efficiency of the exchange algorithm \citep{Benson2021}, one can use an enveloping algorithm akin to an adaptive rejection sampler \citep{Gilks1992}. Still others employ WinBUGs \citep{Lord2012} and the Metropolis-Hastings algorithm \citep{Huang2021} to obtain samples from COM-Poisson-based models.

Much of the previous sampling work is concerned with sampling in the regression case or, in the case of the conjugate prior work \citep{Kadane2006}, places limits on the sampler. Thus, with respect to the previous authors, we develop an alternative approach for sampling from the base COM-Poisson model. To achieve this, we develop \texttt{Stan} code to generate posterior samples through the \texttt{R}-to-\texttt{Stan} interface \texttt{RStan} package \citep{RStan2023}. \texttt{Stan} implements a number of algorithms for performing Bayesian inference including the No U-Turn Sampler \citep{Hoffman2014} and Hamiltonian Monte Carlo \citep{Neal1996}. We leverage the ability to build user-defined posteriors to obtain samples from all of the models in Section~\ref{s:priors}, including the independent Jeffreys' prior-based model, using the No U-Turn Sampler.

\lstset{numbers=left, numberstyle=\tiny, numbersep=5pt, backgroundcolor = \color{lightgray}}
\begin{lstlisting}[float, caption = \texttt{Stan} code for conjugate prior based-models, label = l:conj, language=Stan]
functions {
  real log_Z_terms(int j, real lambda, real nu){
    return(j * log(lambda) - nu * lgamma(j + 1));
  }
}
data {
  int<lower=0> n;   // number of observations 
  int<lower=0> S1;  // sum of X_i's 
  real<lower=0> S2; // sum of log(X_i!) 
  real<lower=0> a;  // hyper-parameter 
  real<lower=0> b;  // hyper-parameter  
  real<lower=0> c;  // hyper-parameter  
}
parameters {
  real<lower=0> lambda;
  real<lower=0> nu;
}
model {
  real logZ[101];
  for (j in 0:100)
    logZ[j+1] = log_Z_terms(j, lambda, nu);
  target += (a + S1 - 1)*log(lambda) - nu*(b + S2) 
		- (c + n)*log_sum_exp(logZ);
}
\end{lstlisting}

Our code is fully available online at \url{https://github.com/markjmeyer/CMP} along with illustrations. We also provide an example of the \texttt{Stan} model for the conjugate prior-based models in Listing~\ref{l:conj}. One issue with any evaluation of COM-Poisson models is how to deal with $Z(\lambda, \nu)$. The infinite sum can be estimated by a finite sum with ``enough'' approximation points; here, we truncate the sum using the first 101 terms. The truncation may appear ``short'' but the additional contribution to the constant by each successive term dies off rapidly. This level of approximation is also based on an upper bound for the truncation error \citep{Minka2003} and is consistent with other existing software packages for COM-Poisson models \citep[Chapter 2]{Sellers2023}. Our online code supplement contains scripts for running all models under consideration.

Starting values for both parameters, $\lambda^{(0)}$ and $\nu^{(0)}$, are selected randomly. Our implementation begins by obtaining two draws, one for each parameter, from a uniform distribution on the interval $(-2,2)$. Because both parameters are bound below by zero, these draws are then exponentiated. Let $u_{\lambda}, u_{\nu} \stackrel{iid}{\sim} U(-2, 2)$ where $U$ denotes the uniform distribution. The starting values are then $\lambda^{(0)} \stackrel{\text{set}}{=} e^{u_{\lambda}}$ and $\nu^{(0)} \stackrel{\text{set}}{=} e^{u_{\nu}}$. On the exponentiated scale, this roughly corresponds to an interval of 0.1353 to 7.3891. For the parameter $\nu$, this interval allows the starting value to be either over-dispersed ($\nu< 1$) or under-dispersed ($\nu>1$).  Thus, the choice of the starting value is not based on any preconceived notion of the dispersion level within the data. The starting value of $\lambda$ comes from the same interval, providing a range of starting values for the generalized rate. We generate four chains, each of length 4000, and retain the last half of each chain for diagnostic checks and inference. This results in 8000 retained posterior samples. To evaluate convergence, we use the potential scale reduction factor or $\hat{R}$ \citep{GelmanRubin1992}. Values of $\hat{R}$ near 1 suggest convergence. This combination of the number of chains and length performed well in our preliminary simulations.

\section{Empirical Study}
\label{s:sim}

Our empirical evaluation of each prior considers three dispersion settings: equi, over, and underdispersion. For the equidispersed setting, we let the true value be $\lambda = 4$ and $\nu=1$. The overdispersed setting has true values of $\lambda = 3$ and $\nu=0.5$ while the underdispersed setting sets $\lambda = 3$ and $\nu=2$. For each setting, we consider sample sizes of $n = 100, 200$, and 500. For each combination of dispersion and $n$, we generate 1000 simulated datasets for evaluation via root mean squared error (rMSE) and leave-one-out cross validation (LOO-CV). All model estimates are based on 8000 retained posterior samples taken over four chains (2000 samples each) after warmups of 2000 samples per chain. Average computation times ranged from 4.538 $s$ to 6.935 $s$ per run across all simulations. Run times varied only slightly by model with the Jeffreys' prior tending to take the longest. Sample size tended to increase computation time as well, although only slightly. All simulations were run on a MacBook Pro laptop with Apple M2 Pro processor and 32GB of RAM.


\begin{table}
	\centering
	\caption{Median estimated hyper-parameters (MAD) for hierarchical reference prior.\label{t:eBayes}}
	\begin{tabular}{llccc}
	  \toprule
	Dispersion & $N$ & $a$ & $b$ & $c$ \\ 
	  \midrule
	Equi & 100 & 0.7092 (0.7452) & 1.9885 (1.3416) & 0.3719 (0.5206) \\ 
	  & 200 & 0.7078 (0.7714) & 2.2443 (1.4729) & 0.4934 (0.7015) \\ 
	  & 500 & 0.5463 (0.6889) & 2.3347 (1.5937) & 0.5495 (0.7949) \\ 
	  \cmidrule{2-5}
	Over & 100 & 0.6946 (0.6058) & 2.0840 (1.3994) & 0.6587 (0.9471) \\ 
	  & 200 & 0.7286 (0.6369) & 2.4391 (1.9196) & 0.7282 (1.0796) \\
	  & 500 & 0.7844 (0.5993) & 2.0779 (1.8943) & 0.2004 (0.2937) \\ 
	  \cmidrule{2-5}
	Under & 100 & 0.5242 (0.5815) & 0.9871 (0.8805) & 0.0461 (0.0642) \\ 
	  & 200 & 0.5879 (0.5701) & 1.0827 (0.8800) & 0.0519 (0.0646) \\ 
	  & 500 & 0.5631 (0.5049) & 0.9440 (0.7545) & 0.0497 (0.0562) \\ 
	   \toprule
	\end{tabular}
\end{table}

The hierarchical reference priors requires we optimize Equation~\ref{eq:kappaBound} for each simulated dataset. In our simulation study, we perform the optimization separately, setting the seed to ensure reproducibility of the datasets. However, the routine is fast enough that it could have been run with the simulation. The median values that maximize Equation~\ref{eq:kappaBound} by setting are in Table~\ref{t:eBayes} along with their median absolute deviations (MAD). The median values of $a$ do not differ much by dispersion type. Median values for $b$ differ slightly more, with underdispersed values tending to be near or under 1. The values for $c$ also tend to be smaller for underdispersed data. Sample size does not appear to impact the values much.

\begin{table}
	\centering
	\caption{Proportion of $\hat{R}$ values under 1.01 by dispersion and parameter for each model and sample size combination.\label{t:rhatProp}}
	\begin{tabular}{lllccc}
	  \toprule
	  &   &   & \multicolumn{3}{c}{$N$} \\
	  \cmidrule{4-6}
	Dispersion & Parameter & Model & 100 & 200 & 500 \\ 
	  \midrule
	Equi & $\lambda$ & Flat & 0.990 & 0.986 & 0.988 \\ 
	  & & Hierarchical & 0.990 & 0.987 & 0.986 \\ 
	  & & Jeffreys & 0.985 & 0.984 & 0.991 \\ 
	  & & Limiting & 0.989 & 0.992 & 0.987 \\ 
	  \cmidrule{3-6}
	& $\nu$ & Flat & 0.990 & 0.984 & 0.988 \\ 
	  & & Hierarchical & 0.992 & 0.986 & 0.985 \\ 
	  & & Jeffreys & 0.986 & 0.982 & 0.992 \\ 
	  & & Limiting & 0.988 & 0.992 & 0.983 \\ 
	  \cmidrule{2-6}
	Over & $\lambda$ & Flat & 0.984 & 0.990 & 0.987 \\ 
	  & & Hierarchical & 0.987 & 0.988 & 0.988 \\ 
	  & & Jeffreys & 0.984 & 0.982 & 0.989 \\ 
	  & & Limiting & 0.988 & 0.984 & 0.982 \\ 
	  \cmidrule{3-6}
	& $\nu$ & Flat & 0.981 & 0.990 & 0.985 \\ 
	  & & Hierarchical & 0.986 & 0.986 & 0.990 \\ 
	  & & Jeffreys & 0.984 & 0.985 & 0.988 \\ 
	  & & Limiting & 0.990 & 0.983 & 0.982 \\ 
	  \cmidrule{2-6}
	Under & $\lambda$ & Flat & 0.998 & 0.998 & 0.998 \\ 
	  & & Hierarchical & 0.996 & 0.997 & 0.995 \\ 
	  & & Jeffreys & 0.998 & 0.998 & 0.996 \\ 
	  & & Limiting & 0.999 & 0.998 & 0.998 \\ 
	  \cmidrule{3-6}
	& $\nu$ & Flat & 0.998 & 0.997 & 0.998 \\ 
	  & & Hierarchical & 0.997 & 0.995 & 0.993 \\ 
	  & & Jeffreys & 0.998 & 0.998 & 0.997 \\ 
	  & & Limiting & 1.000 & 0.998 & 0.999 \\ 
	   \toprule
	\end{tabular}
\end{table}

%

We use $\hat{R}$ to assess convergence across chains. Table~\ref{t:rhatProp} contains the proportion of $\hat{R}$ values that fall below 1.01 across all 1000 simulated datasets for each parameter under every combination of model, dispersion, and sample size. Across all simulated datasets, we observe $\hat{R}$ values to all be smaller than 1.0205. Thus, we judge the chains to have converged for all models. We also note that no simulation setting produced divergent transitions in the No U-Turn Sampler.


\begin{figure}
	\centering
	\includegraphics[scale = 0.35]{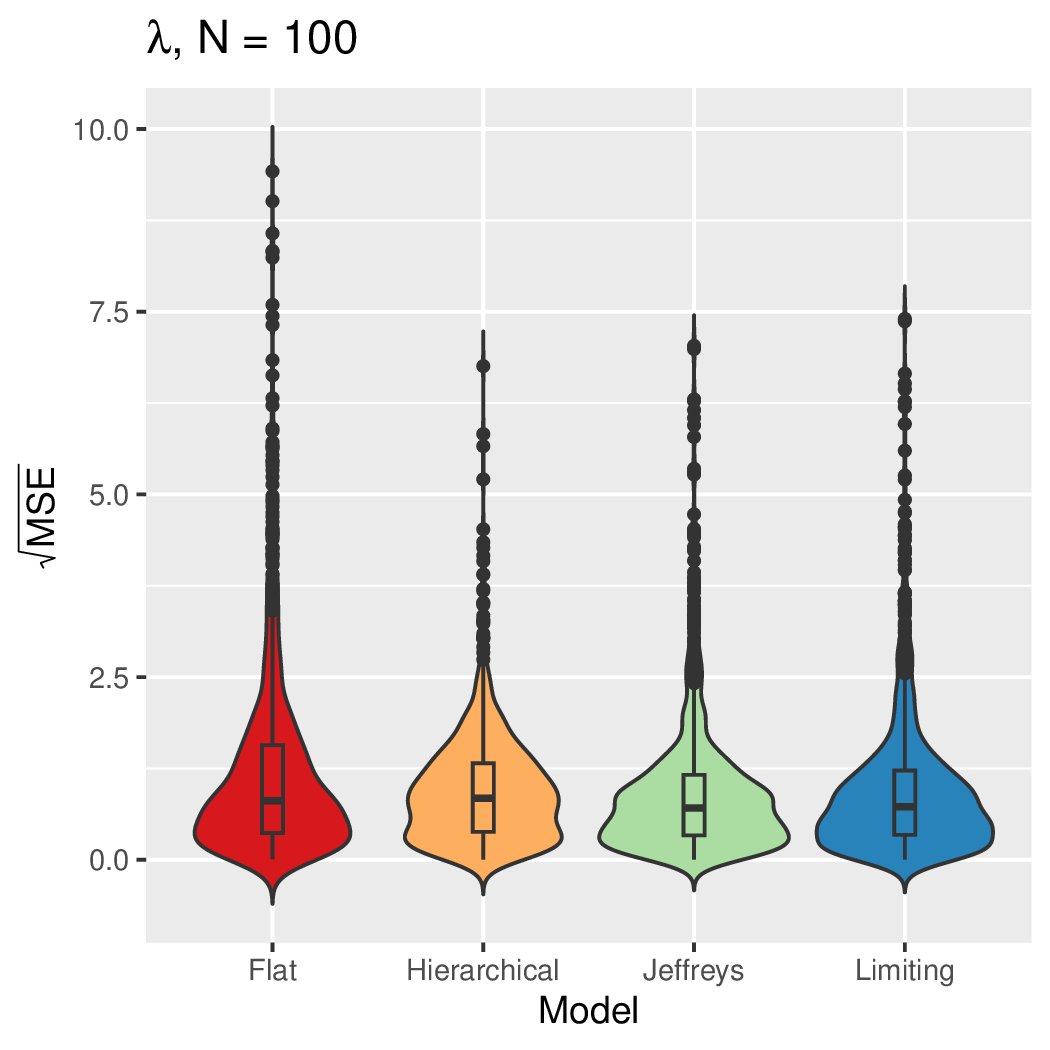}
	\includegraphics[scale = 0.35]{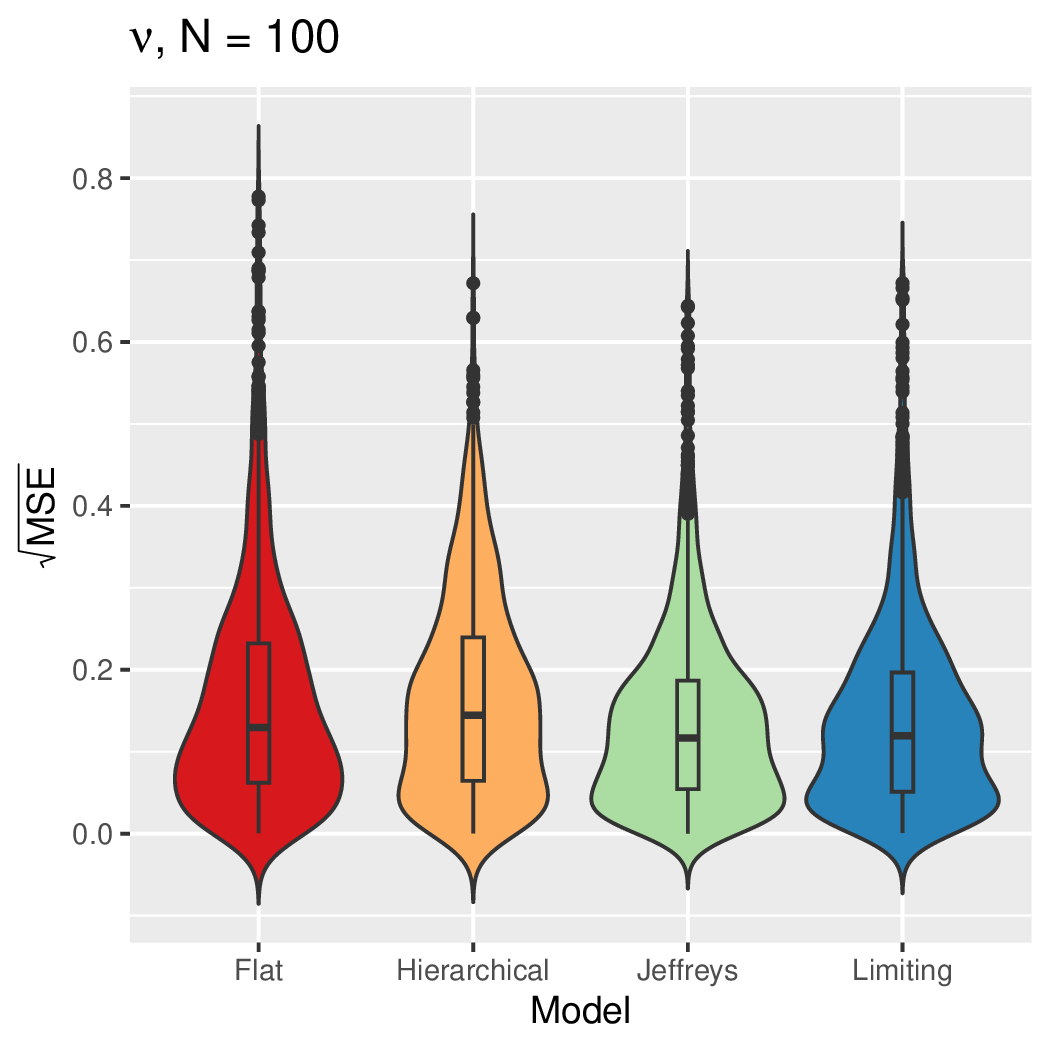}
	\includegraphics[scale = 0.35]{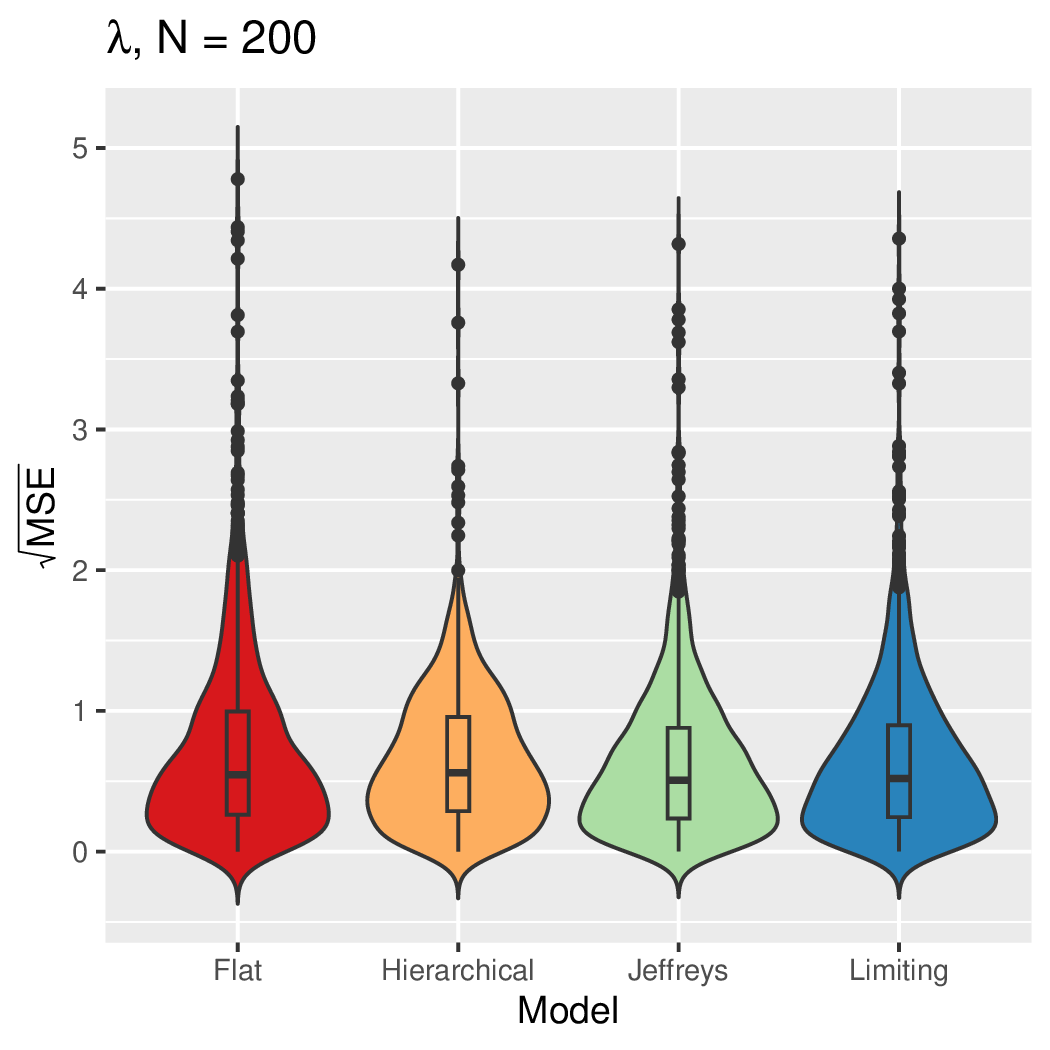}
	\includegraphics[scale = 0.35]{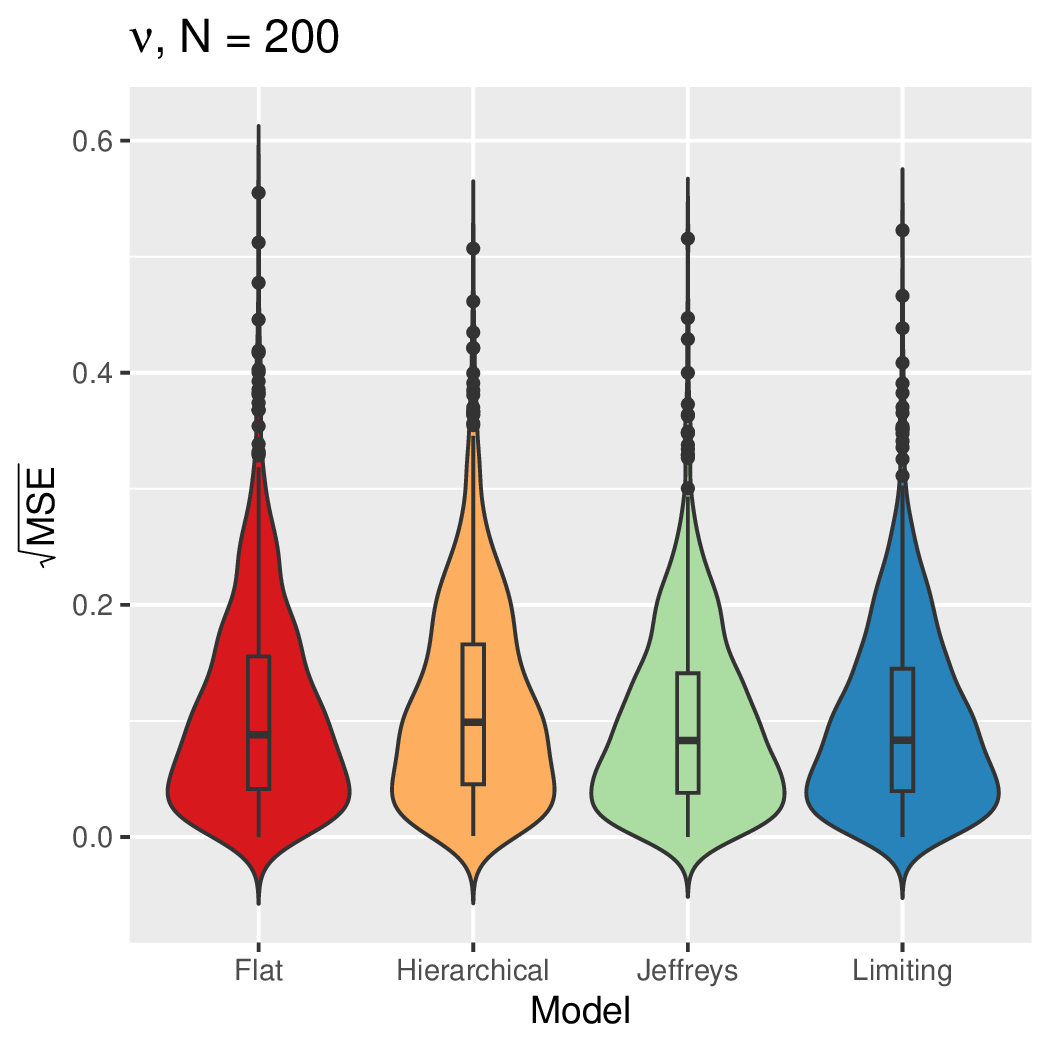}
	\includegraphics[scale = 0.35]{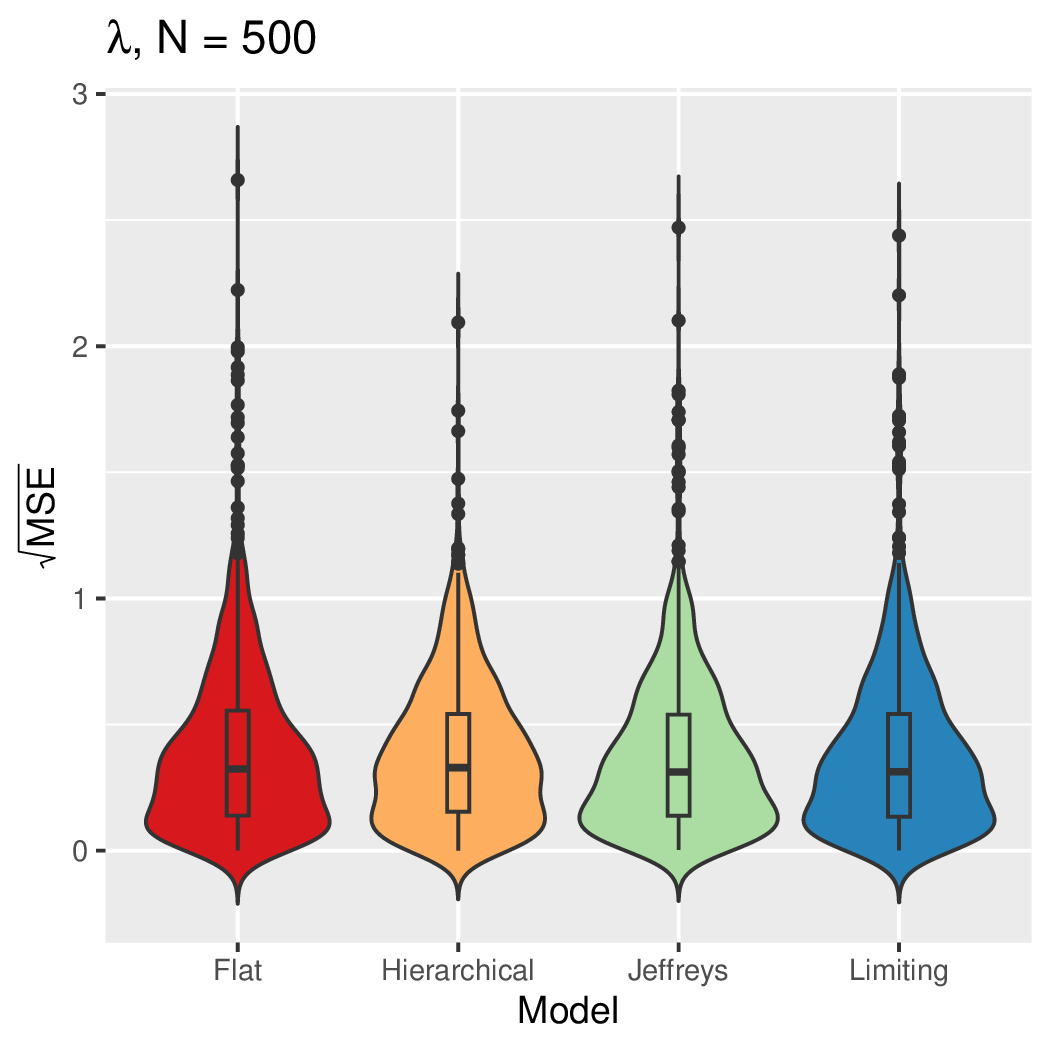}
	\includegraphics[scale = 0.35]{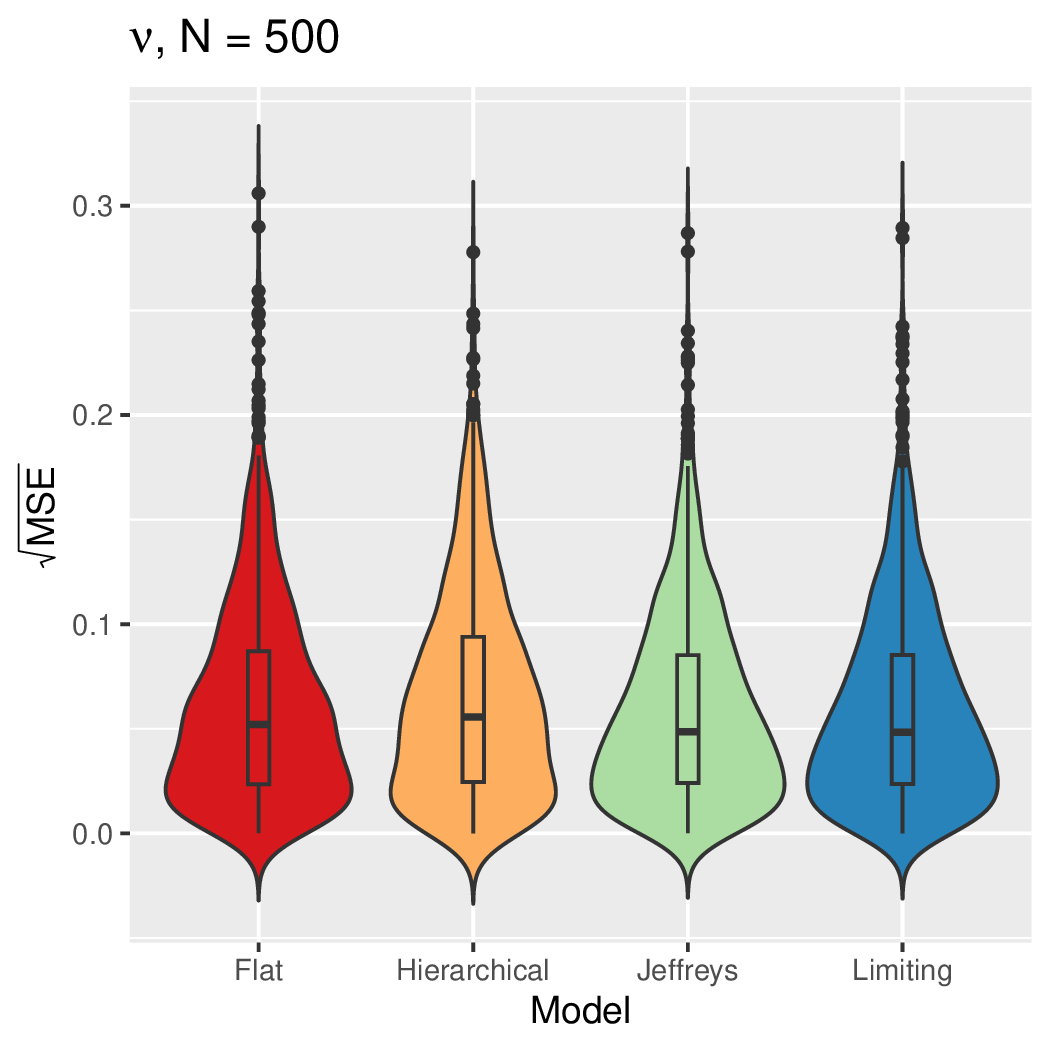}
	\caption{Root mean squared error ($\sqrt{\text{MSE}}$) of $\lambda$ and $\nu$ for the equidispersed setting, by model and sample size.\label{f:prmse}}
\end{figure}

\begin{figure}
	\centering
	\includegraphics[scale = 0.35]{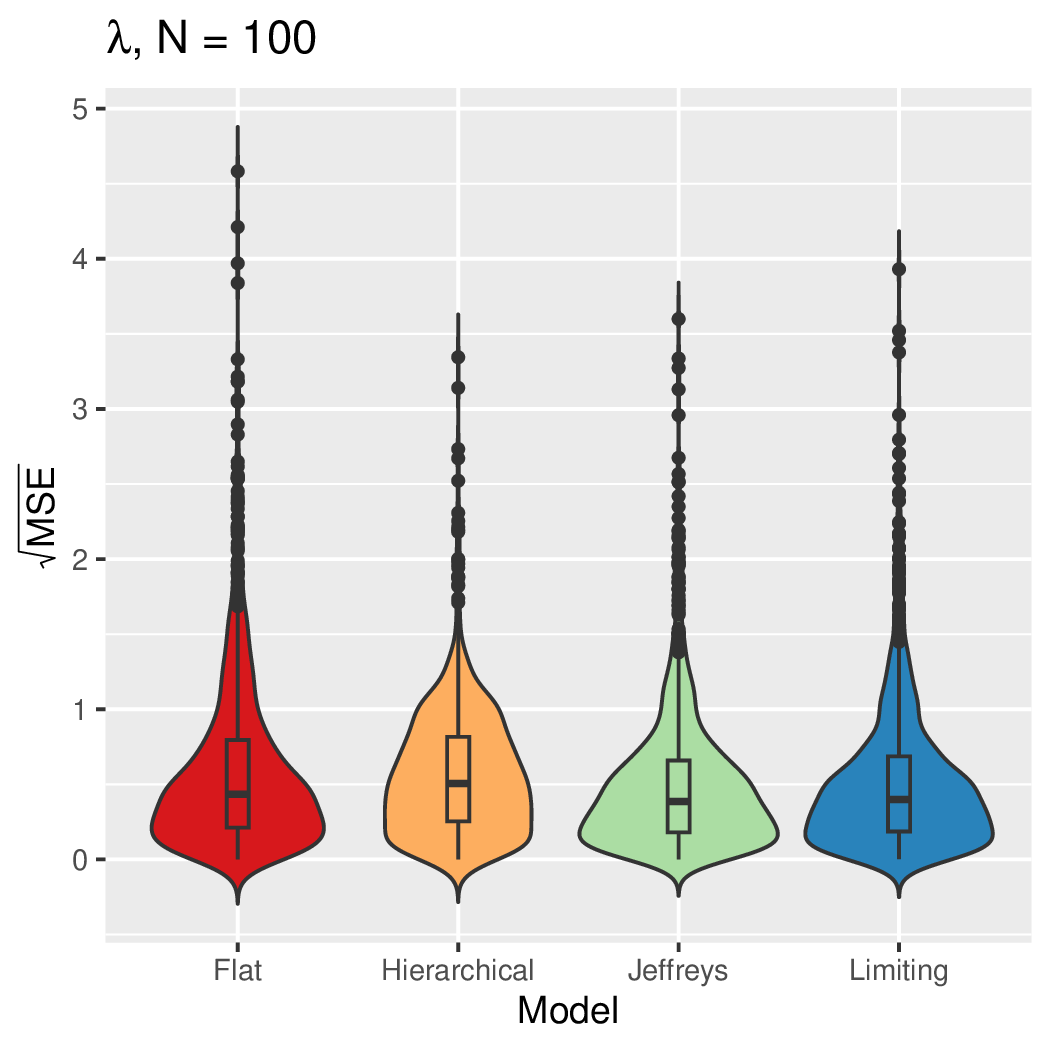}
	\includegraphics[scale = 0.35]{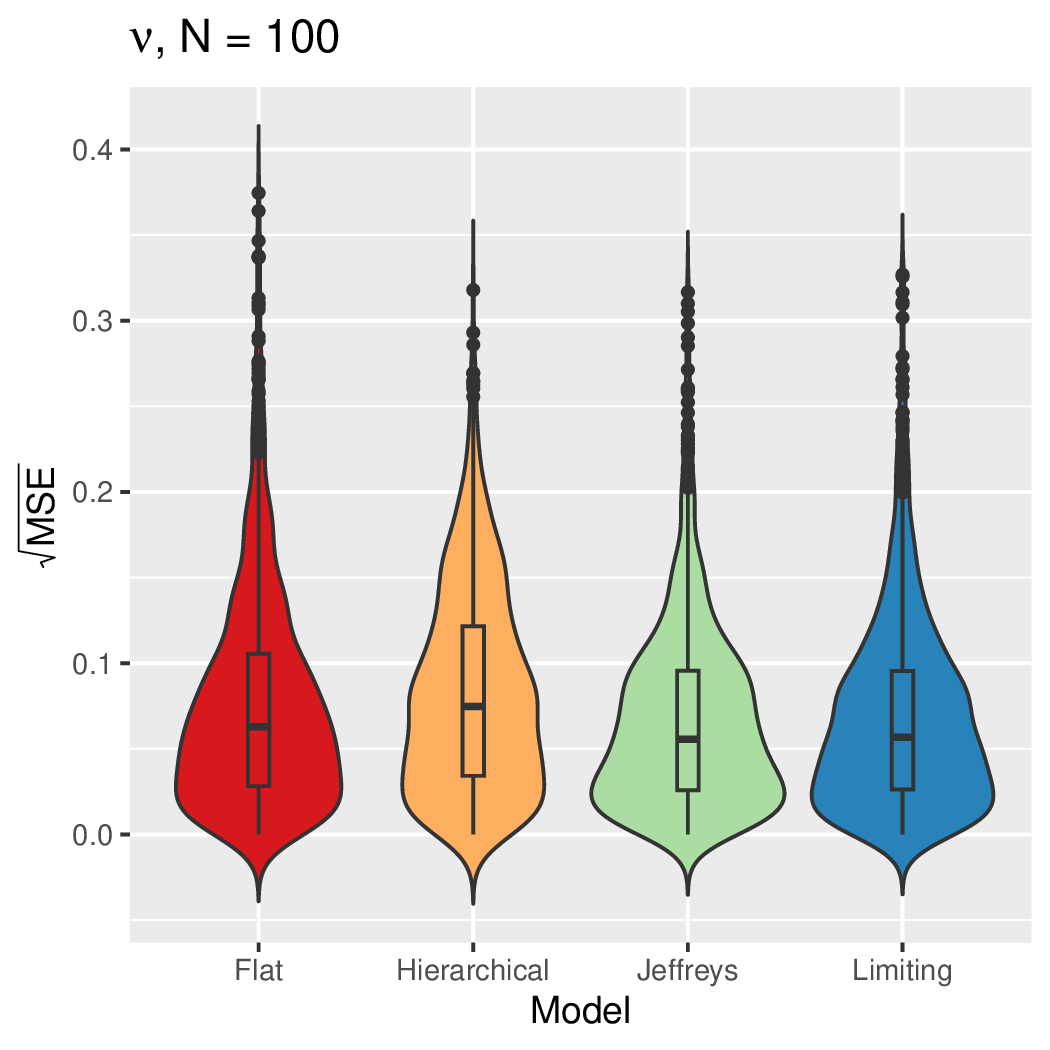}
	\includegraphics[scale = 0.35]{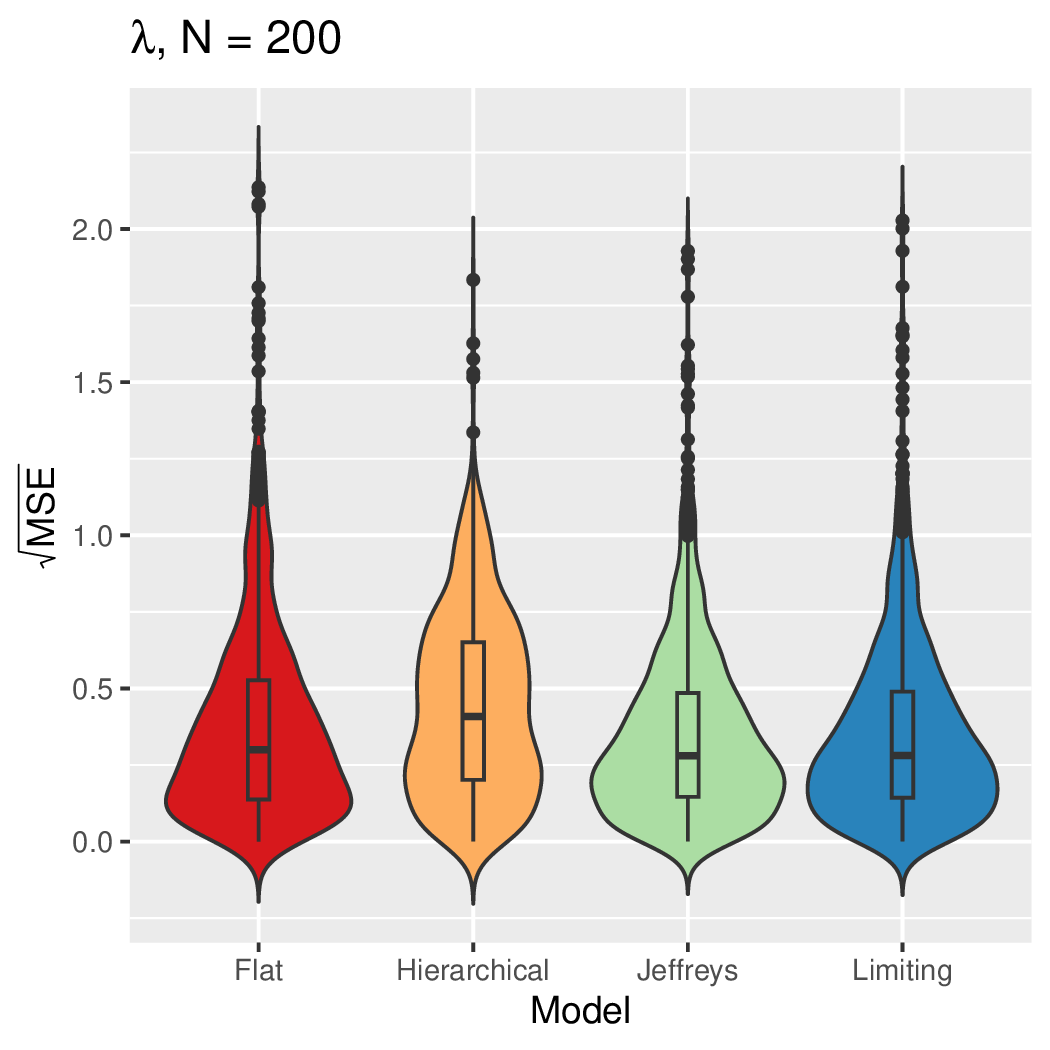}
	\includegraphics[scale = 0.35]{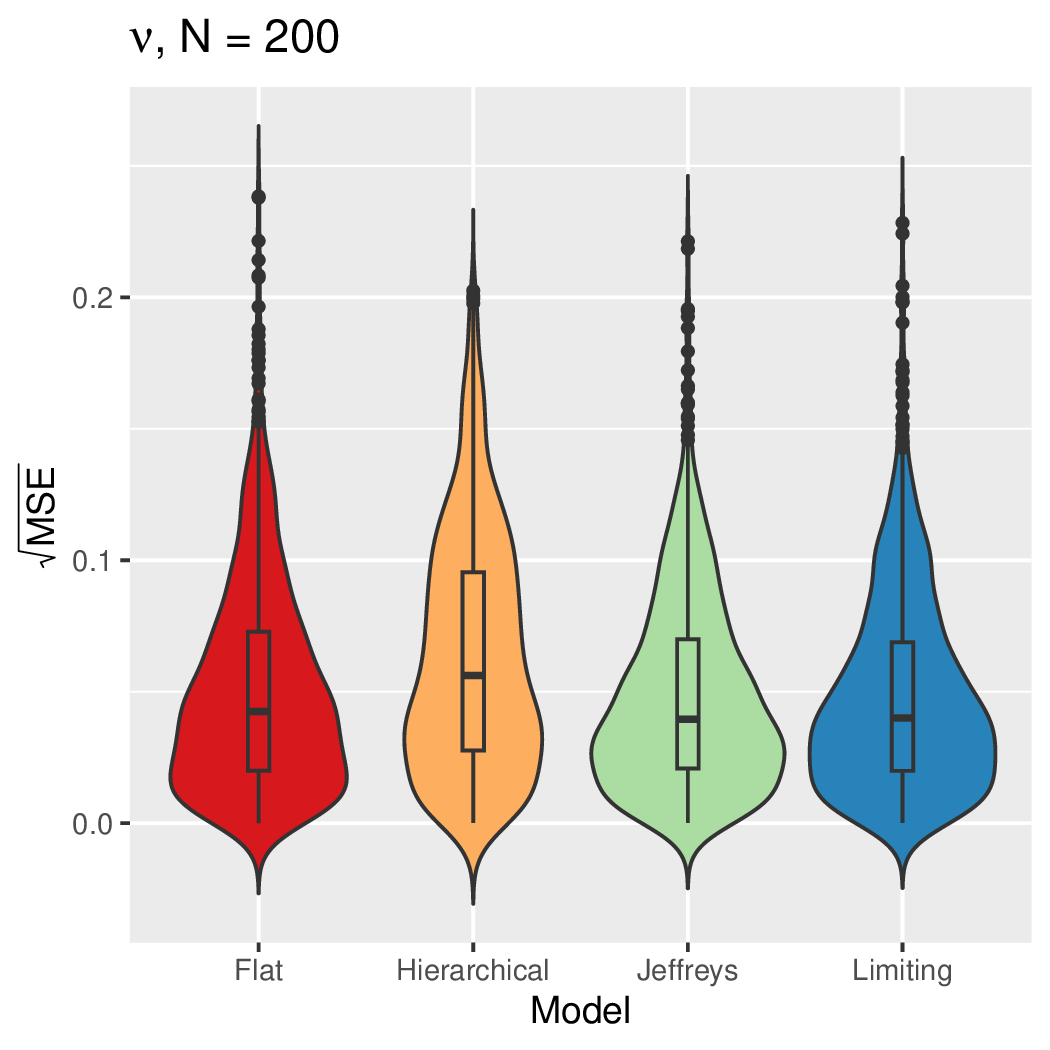}
	\includegraphics[scale = 0.35]{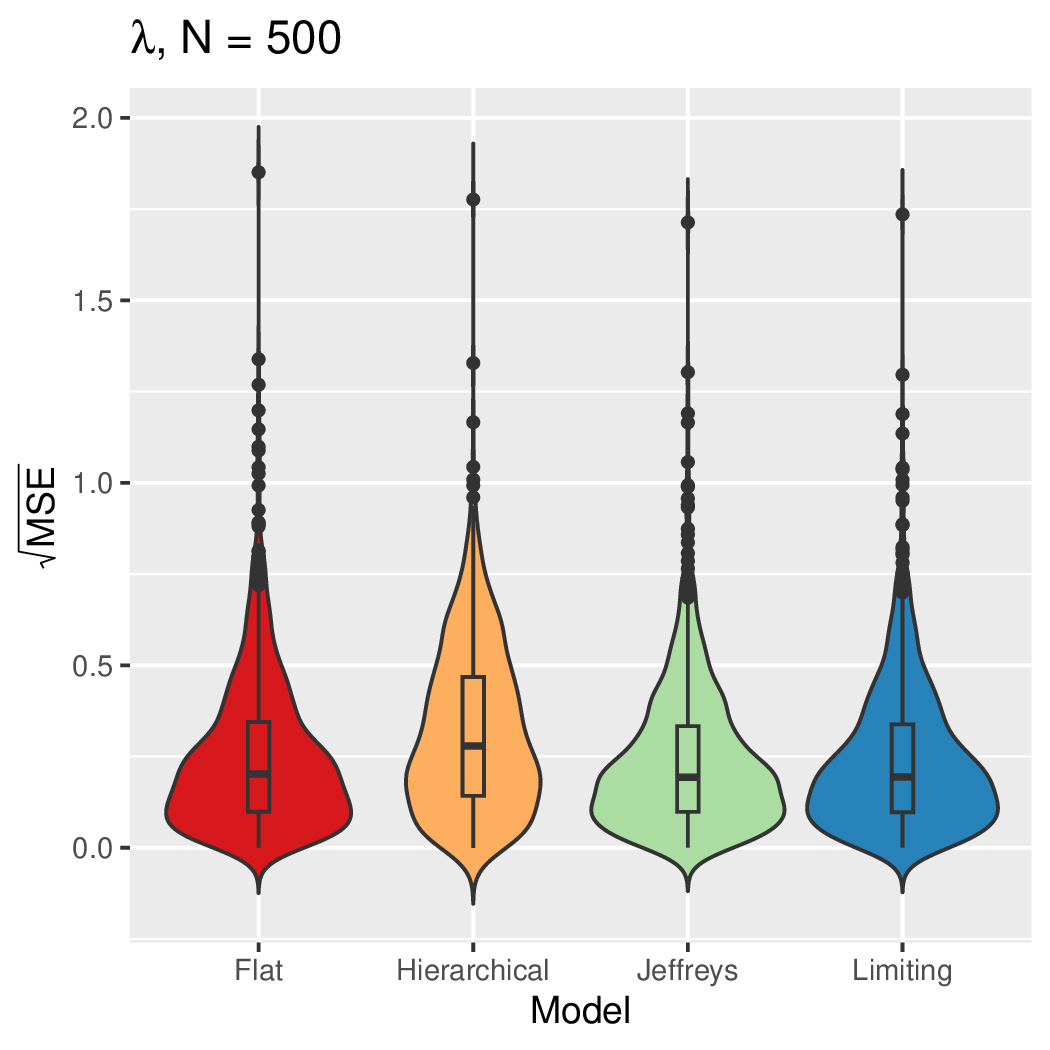}
	\includegraphics[scale = 0.35]{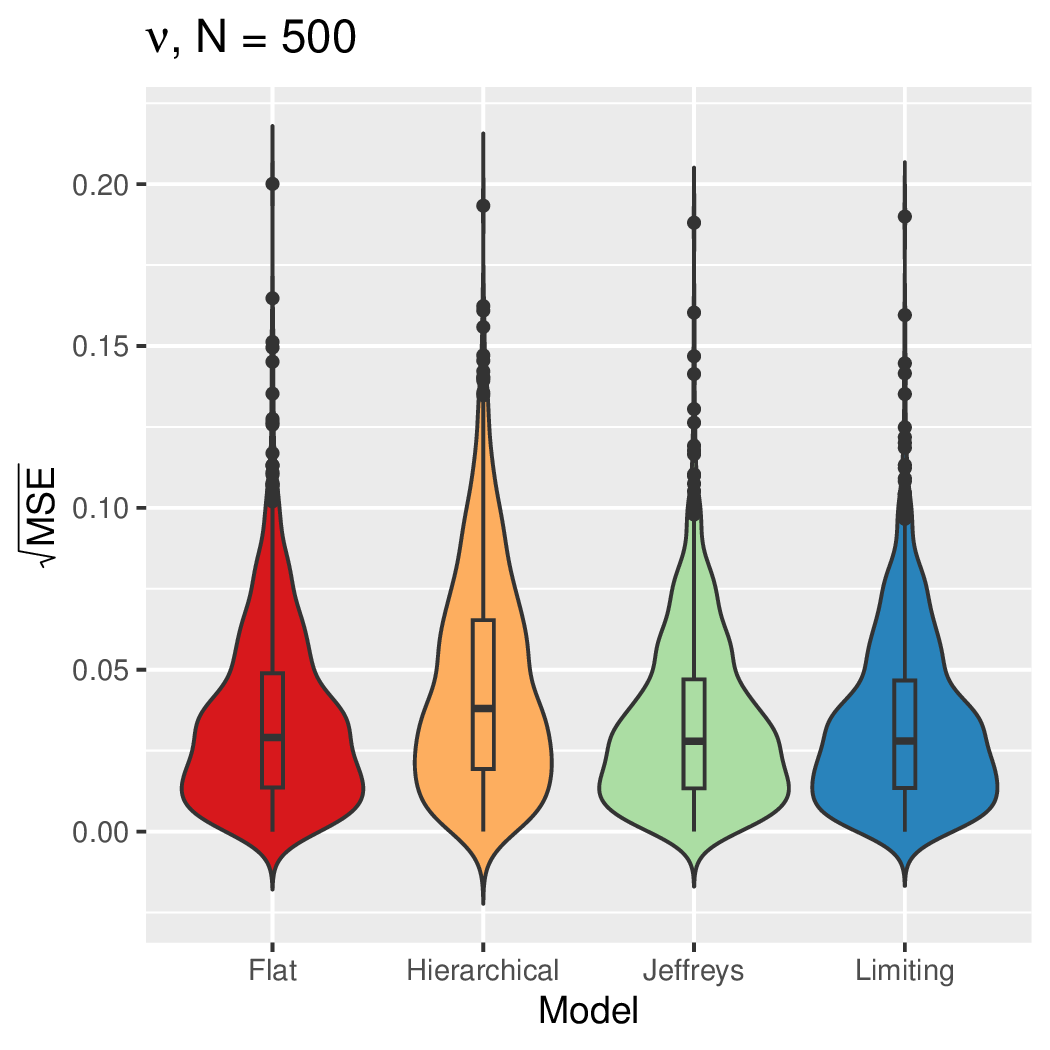}
	\caption{Root mean squared error ($\sqrt{\text{MSE}}$) of $\lambda$ and $\nu$ for the overdispersed setting, by model and sample size.\label{f:ormse}}
\end{figure}

\begin{figure}
	\centering
	\includegraphics[scale = 0.35]{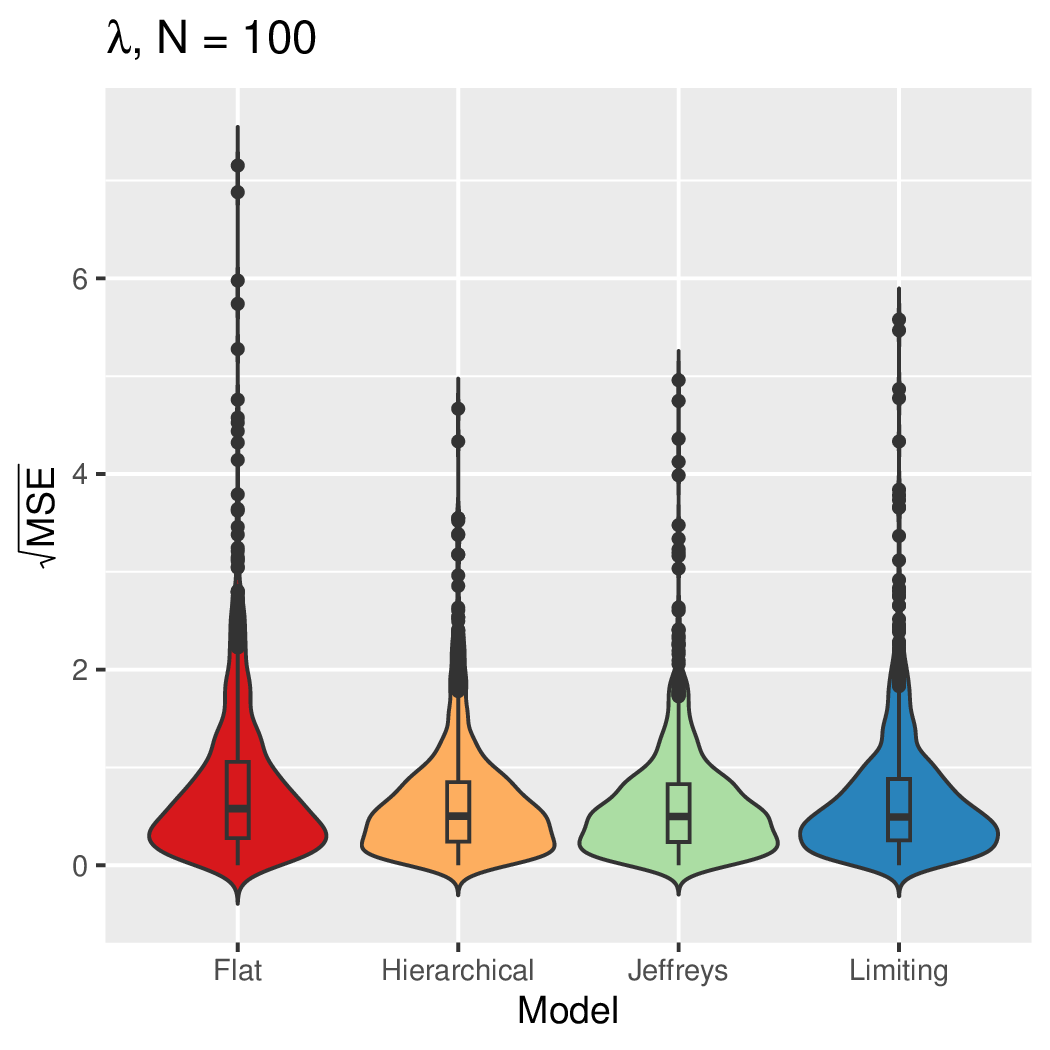}
	\includegraphics[scale = 0.35]{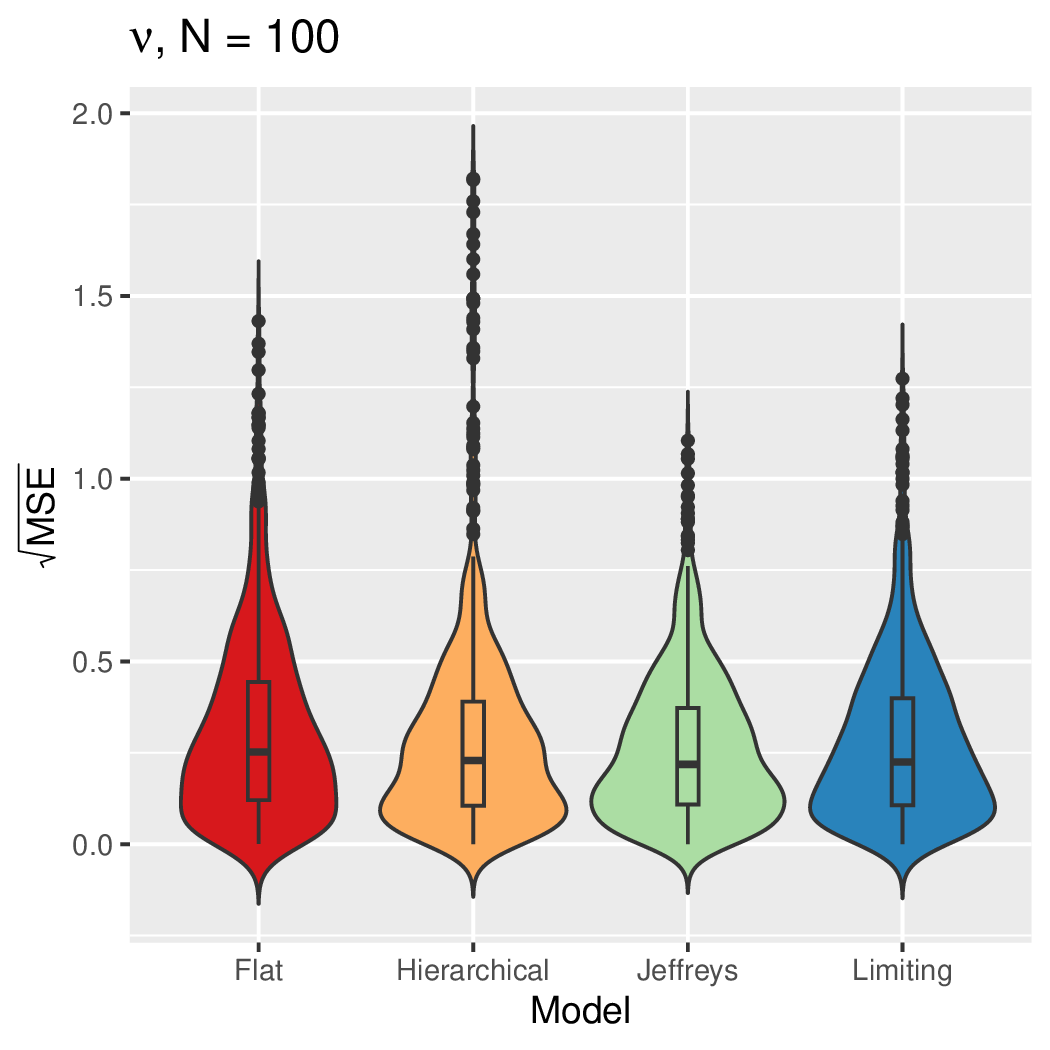}
	\includegraphics[scale = 0.35]{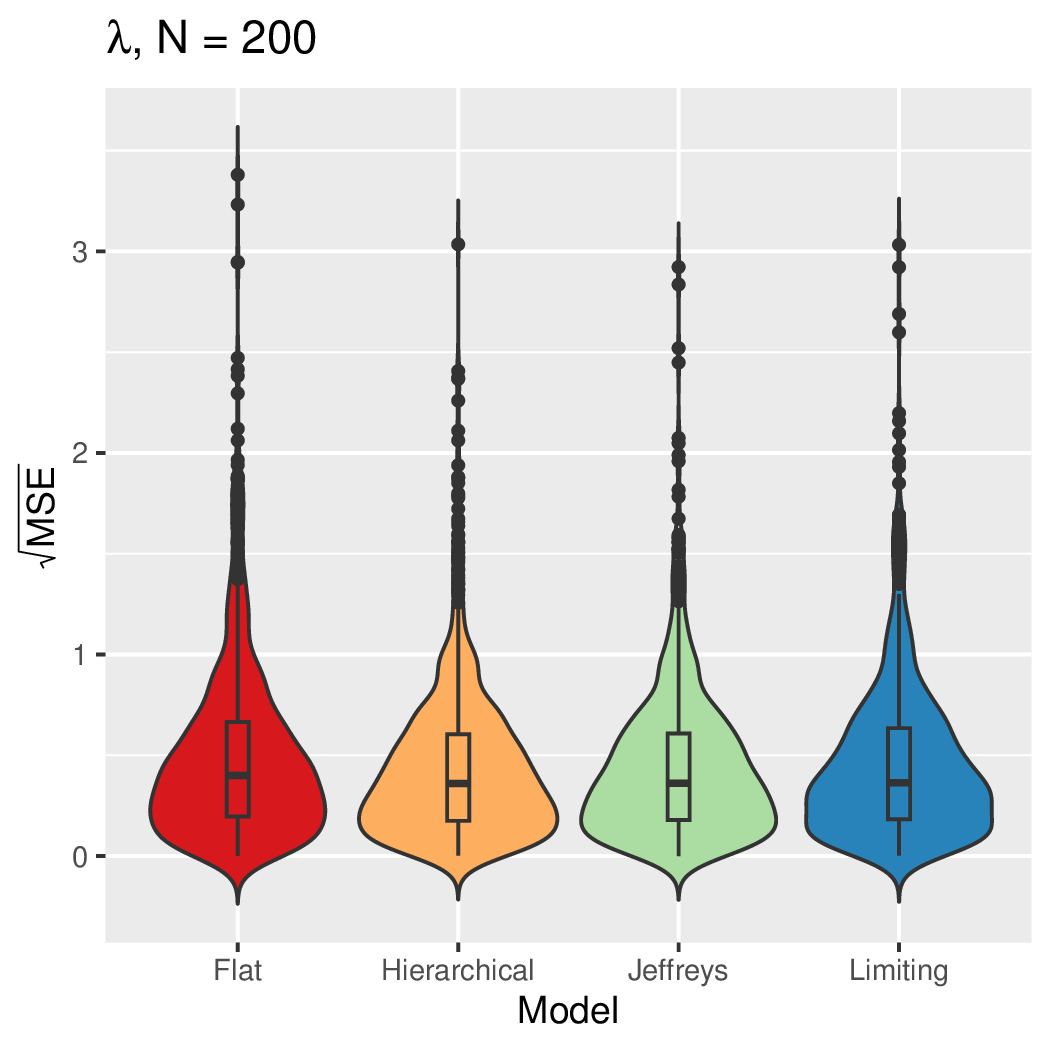}
	\includegraphics[scale = 0.35]{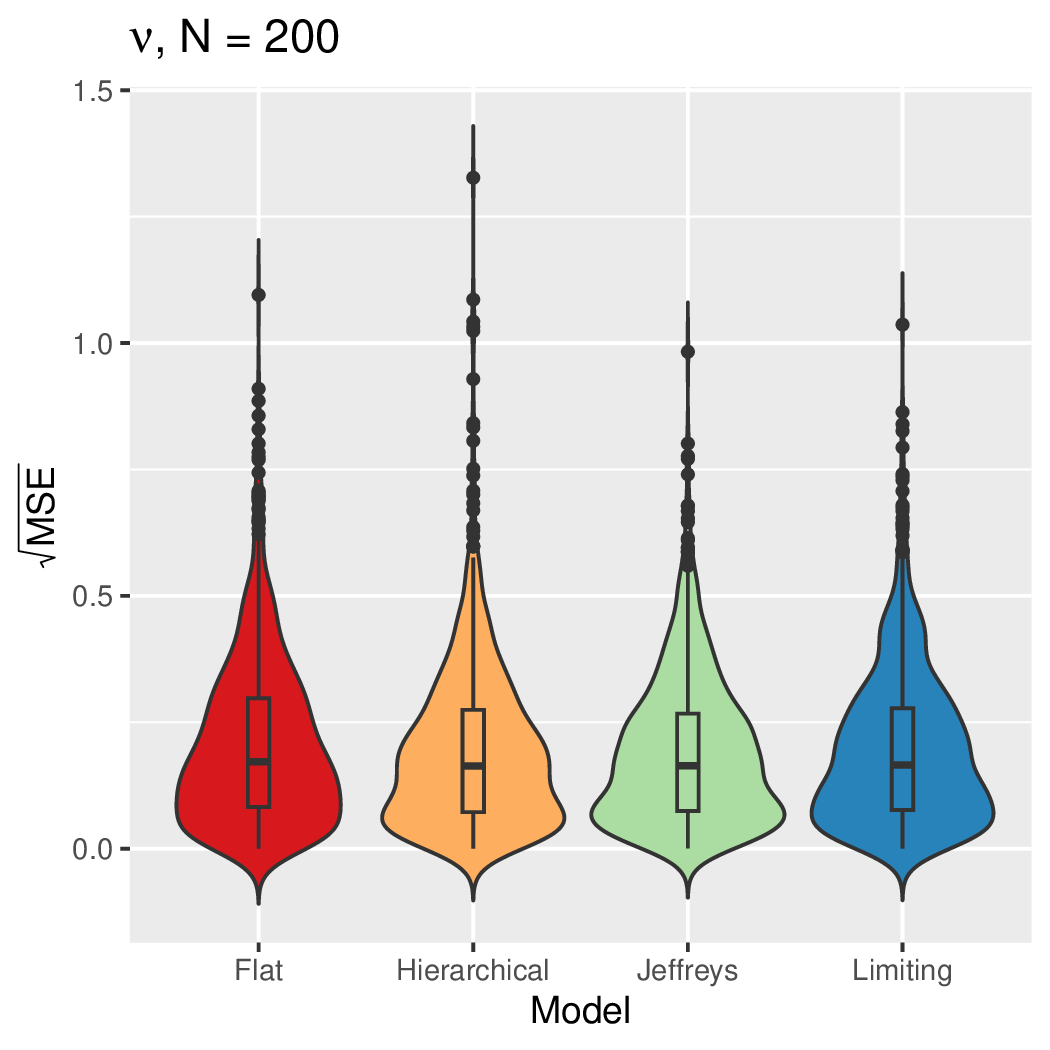}
	\includegraphics[scale = 0.35]{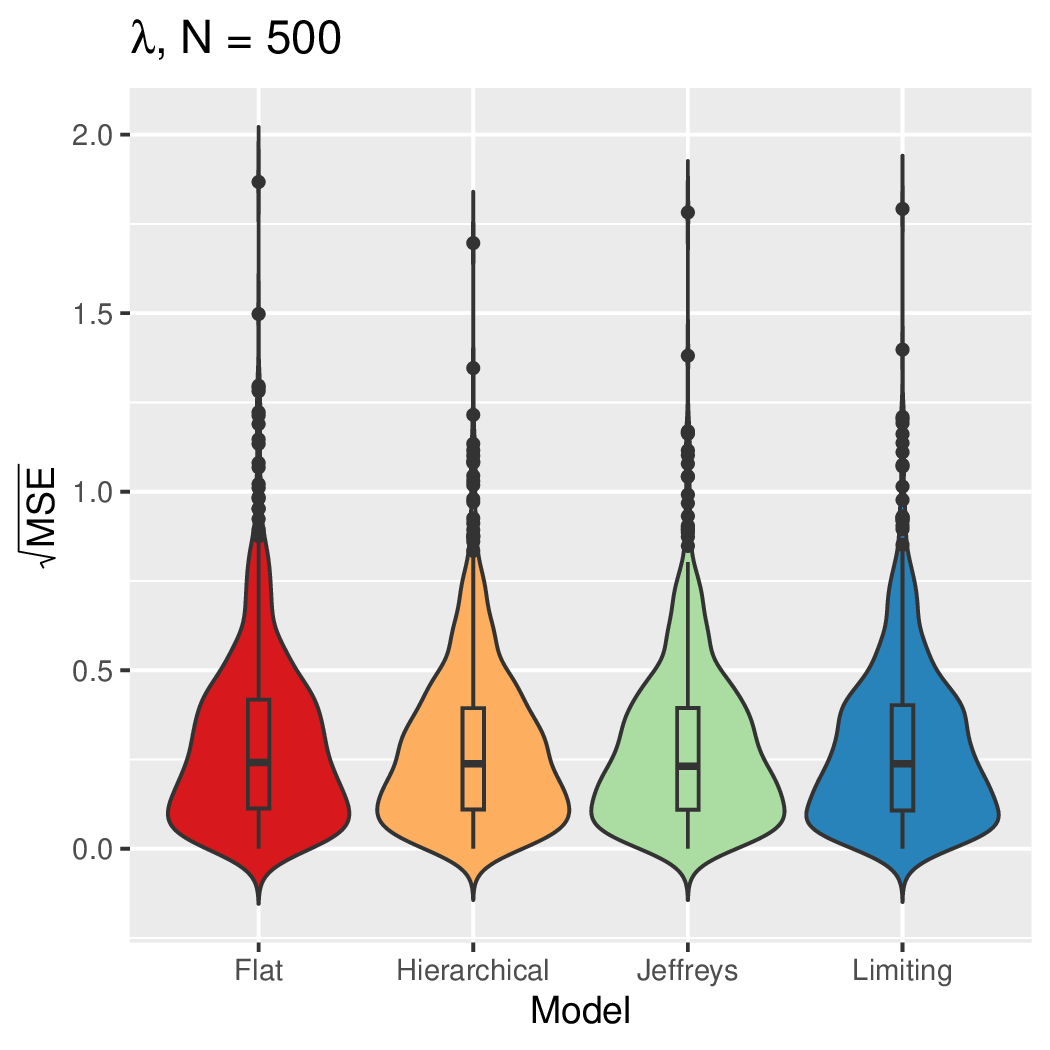}
	\includegraphics[scale = 0.35]{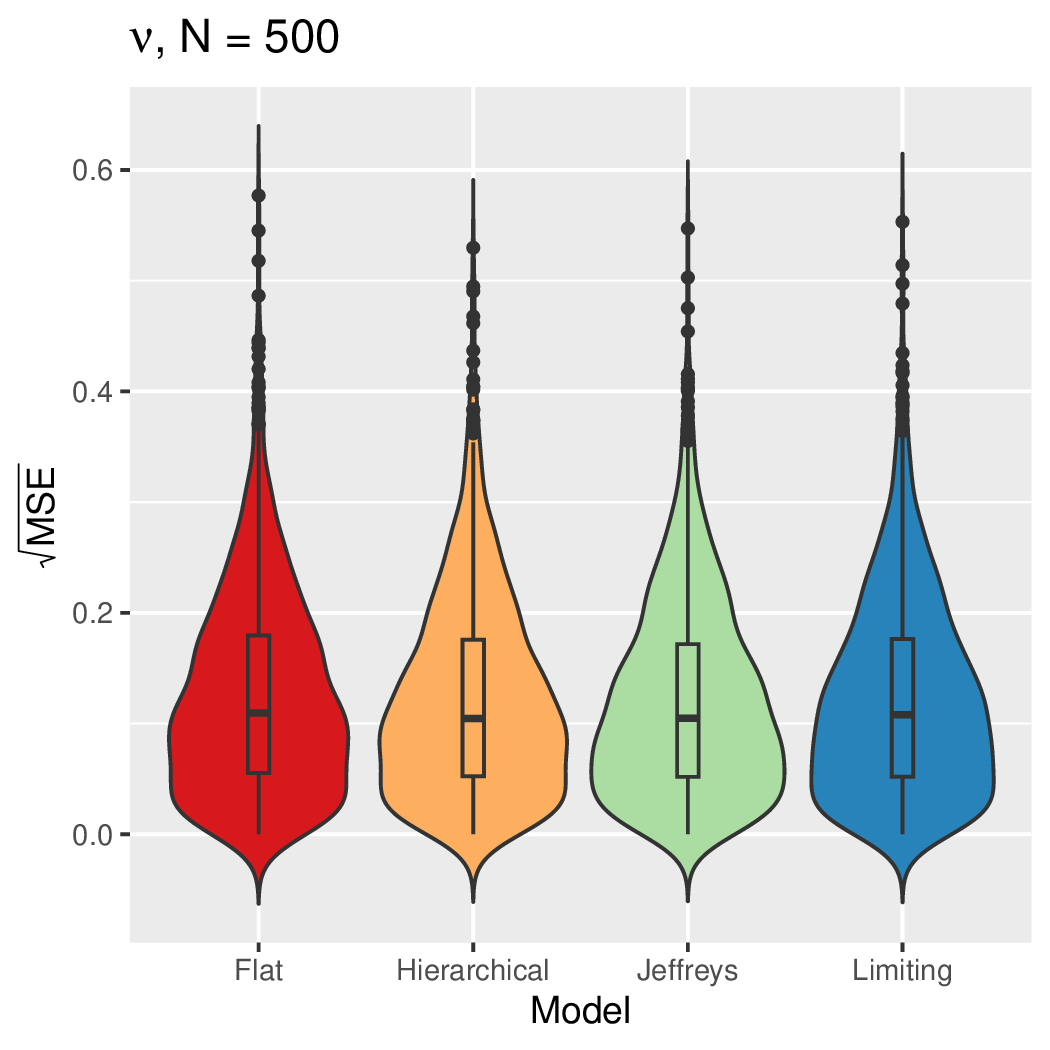}
	\caption{Root mean squared error ($\sqrt{\text{MSE}}$) of $\lambda$ and $\nu$ for the underdispersed setting, by model and sample size.\label{f:urmse}}
\end{figure}

Our comparison metrics between models are rMSE and LOO-CV. Violin plots for the rMSE values are in Figures~\ref{f:prmse} to ~\ref{f:urmse}. In each figure, the left column contains rMSE values for $\lambda$ while the right contains values for $\nu$. The rows correspond to sample sizes with top row containing values when $N = 100$, second row containing values when $N = 200$, and the bottom row containing values when $N = 500$. The red violin plots are for rMSE values from the model using the join flat prior, orange for the model using the hierarchical reference prior, green for the model using the independent Jefffreys' prior, and blue for the model using the joint limiting prior. Overall, we do not see major differences between the prior models in terms of rMSE. Sample size appears to be the largest driver in reducing rMSE. While all models exhibit some large values, the flat prior tends to have the widest overall spread. Across dispersion, rMSE tends to be largest for $\lambda$ when the data is equidispersed while it is largest for $\nu$ when the data is underdispersed.

\begin{figure}
	\centering
	\includegraphics[scale = 0.35]{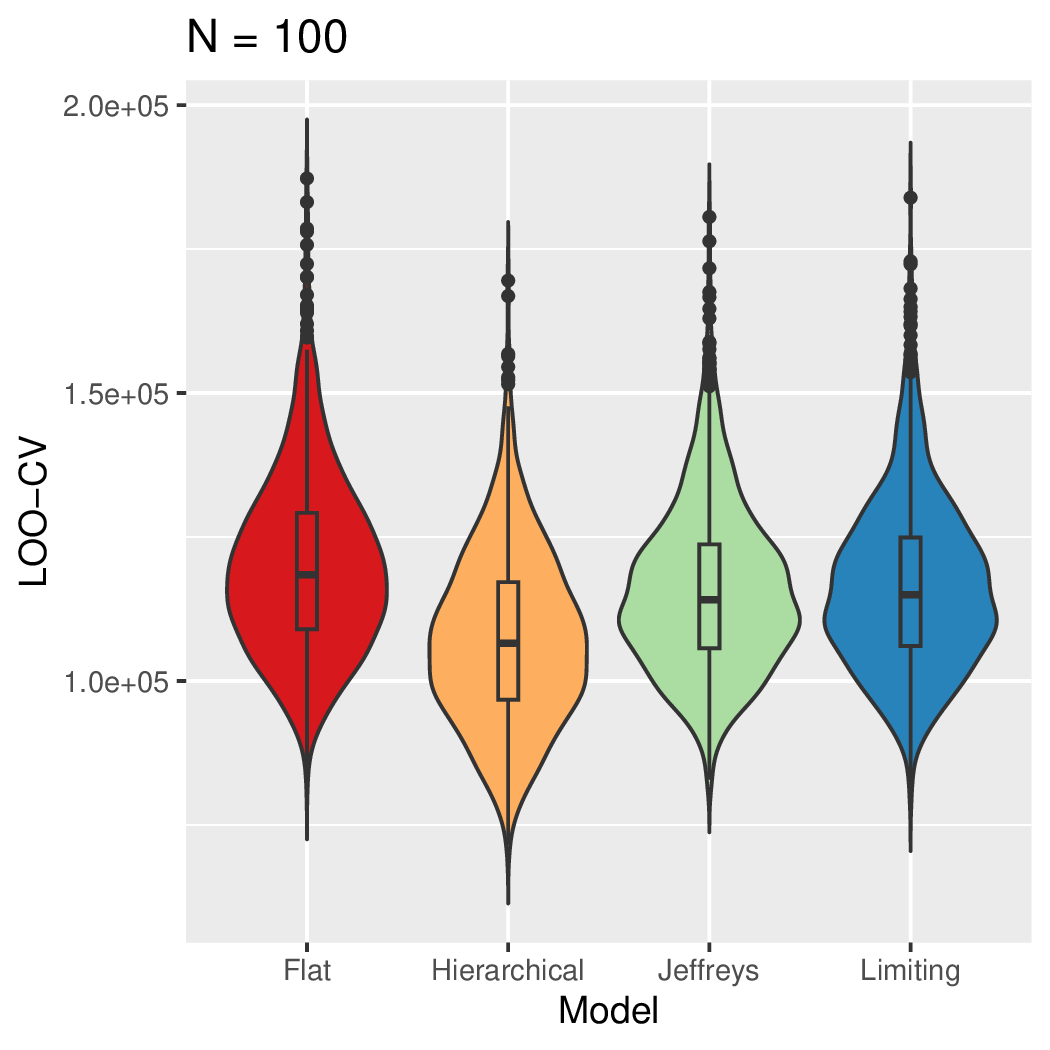}
	\includegraphics[scale = 0.35]{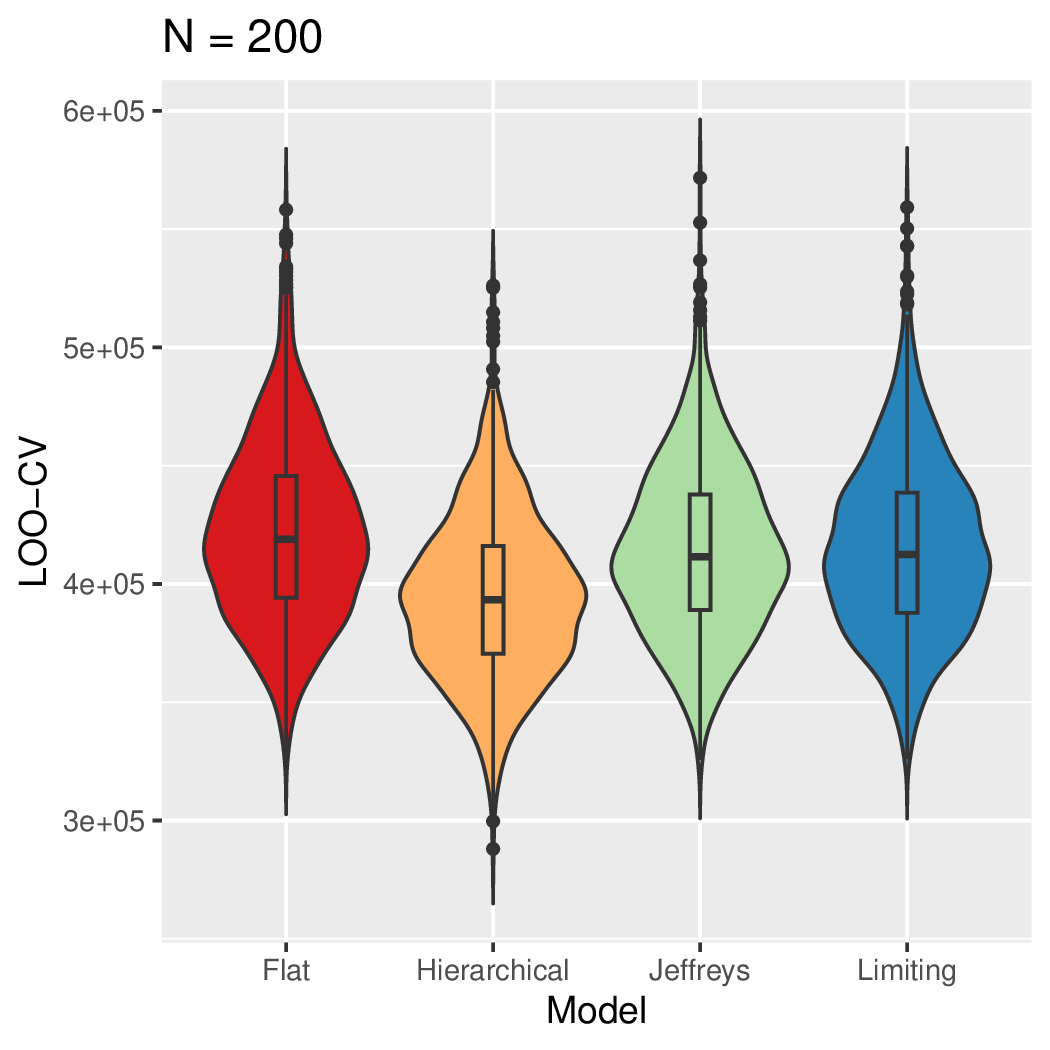}
	\includegraphics[scale = 0.35]{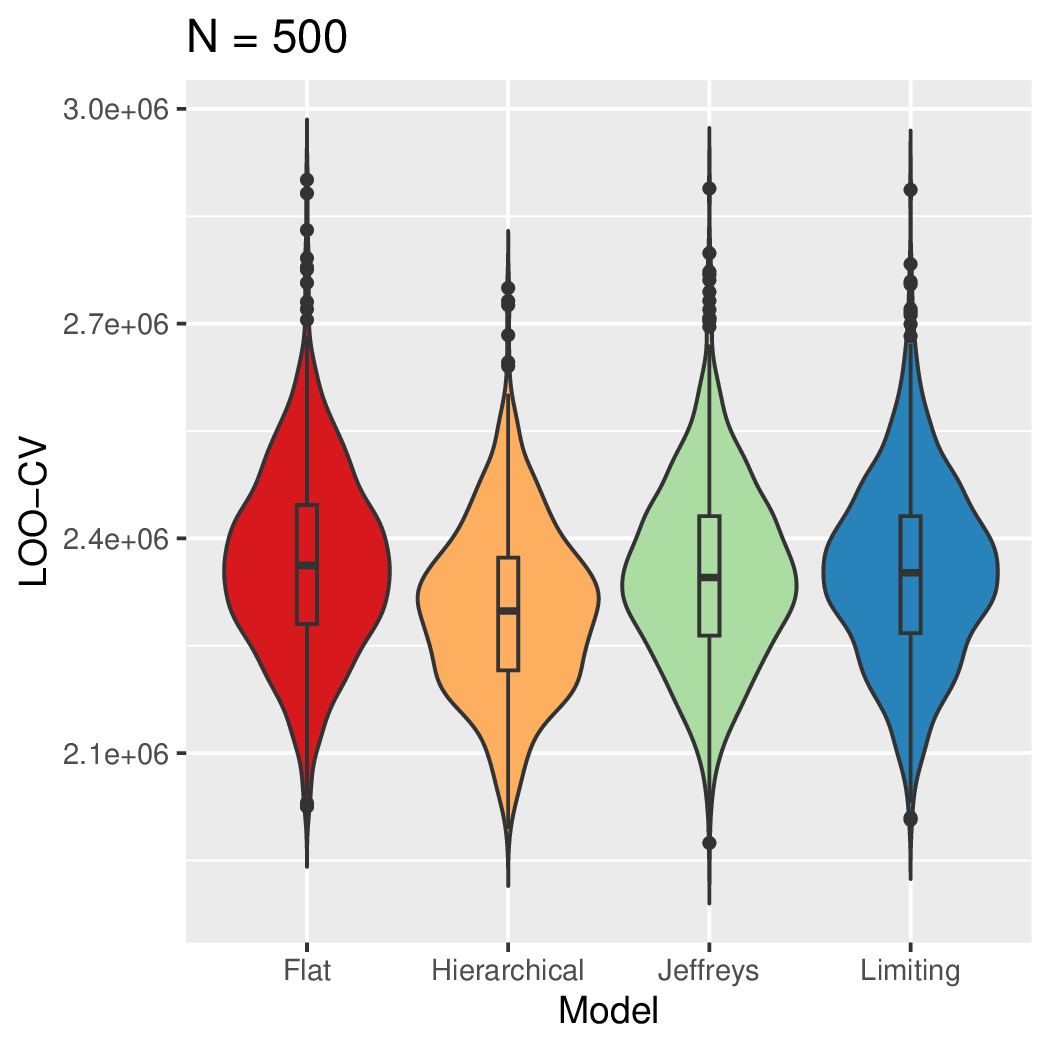}
	\caption{Leave-one-out Cross Validation (LOO-CV) for the equidispersed setting, by model and sample size.\label{f:ploos}}
\end{figure}

\begin{figure}
	\centering
	\includegraphics[scale = 0.35]{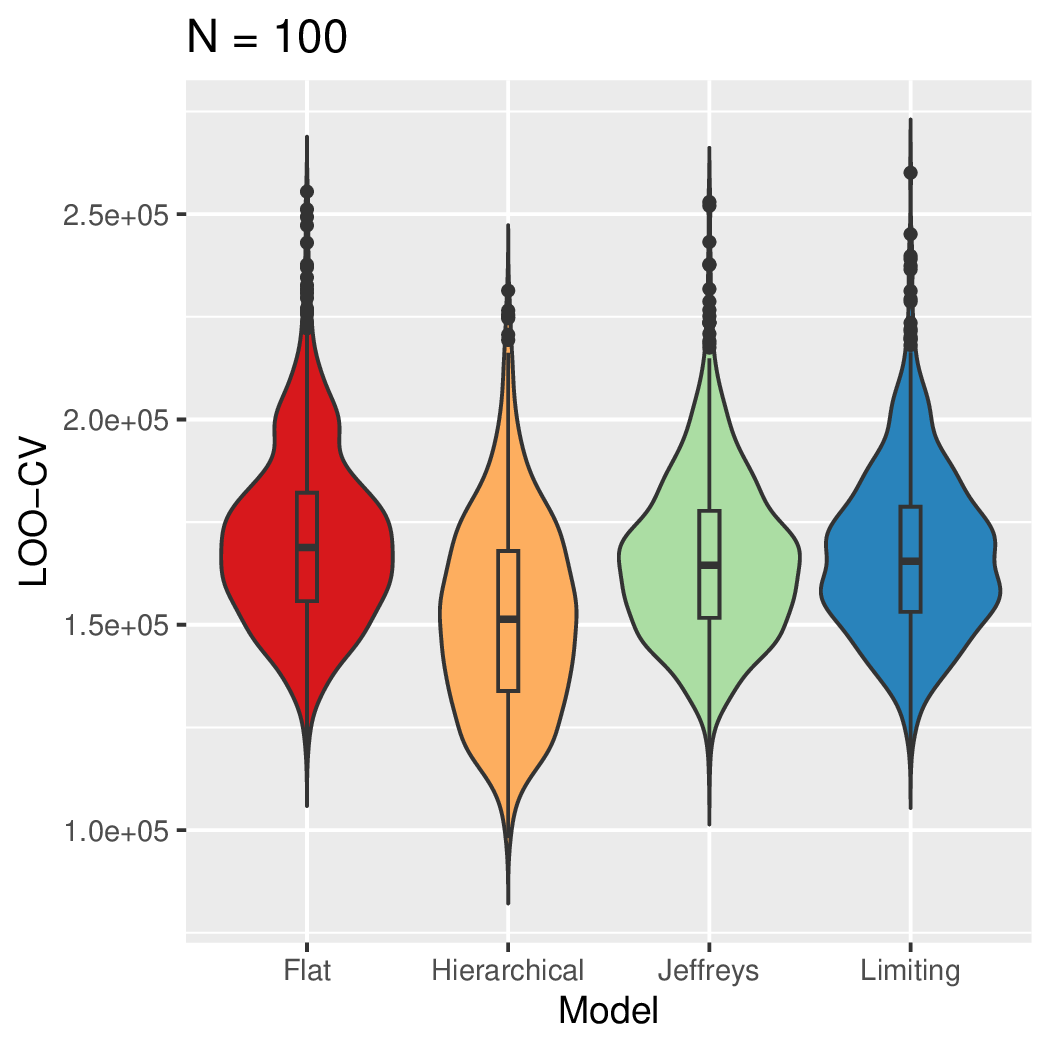}
	\includegraphics[scale = 0.35]{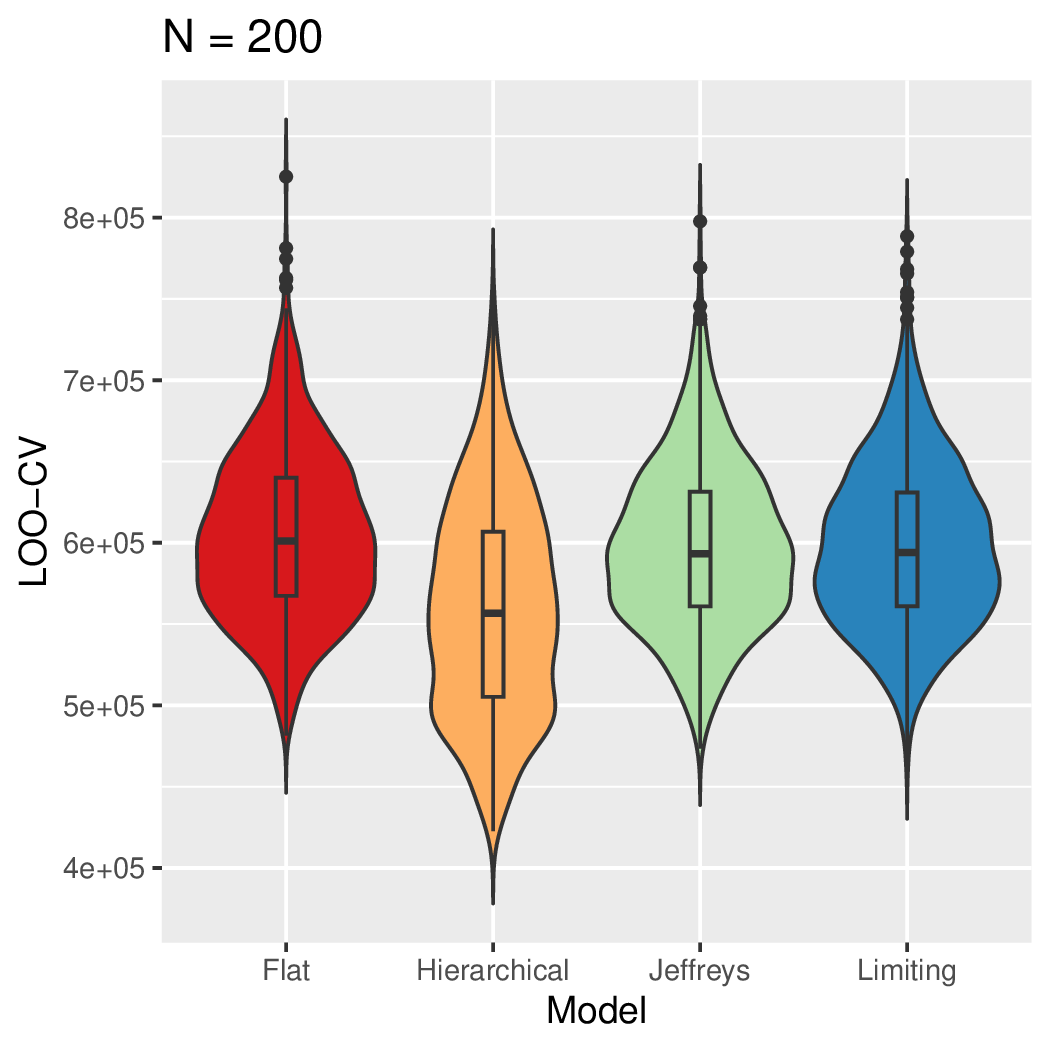}
	\includegraphics[scale = 0.35]{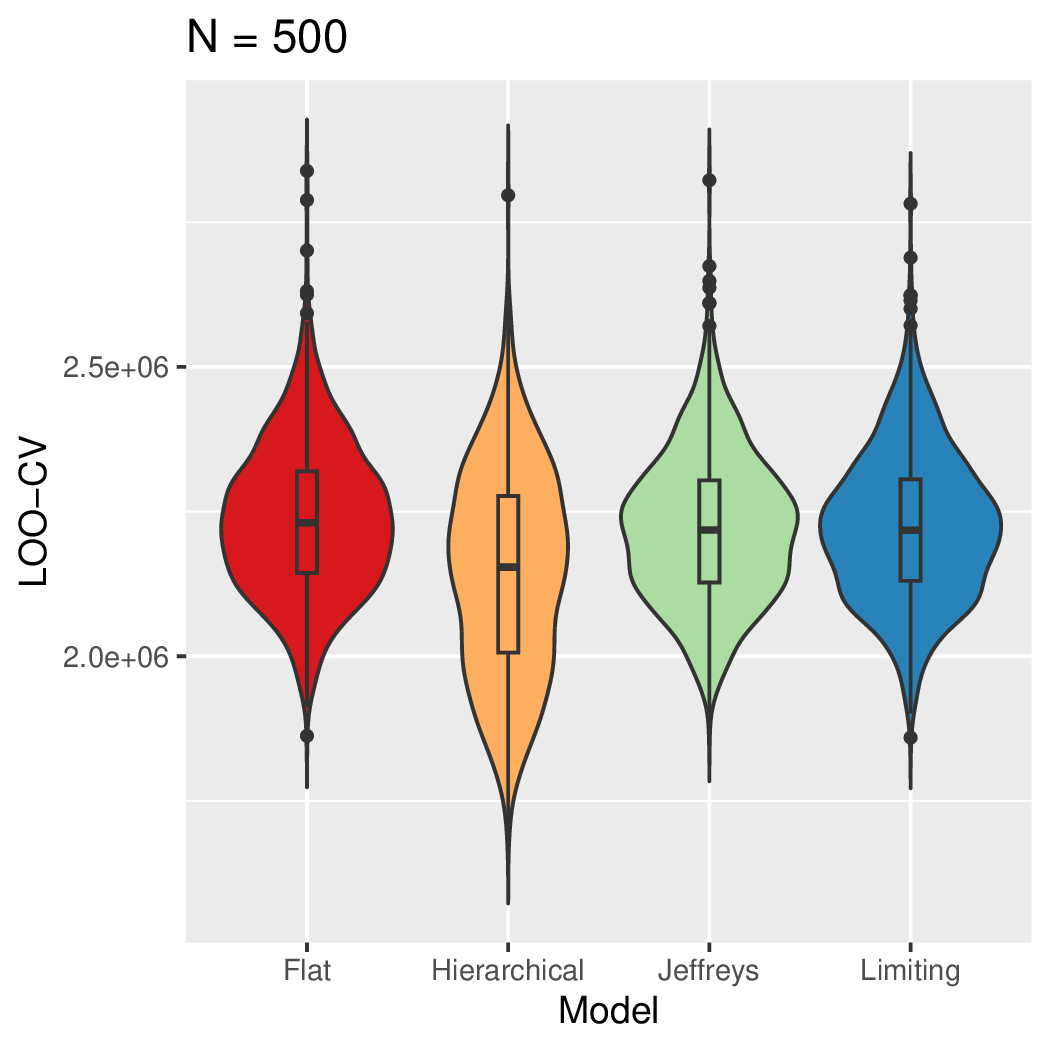}
	\caption{Leave-one-out Cross Validation (LOO-CV) for the overdispersed setting, by model and sample size.\label{f:oloos}}
\end{figure}

\begin{figure}
	\centering
	\includegraphics[scale = 0.35]{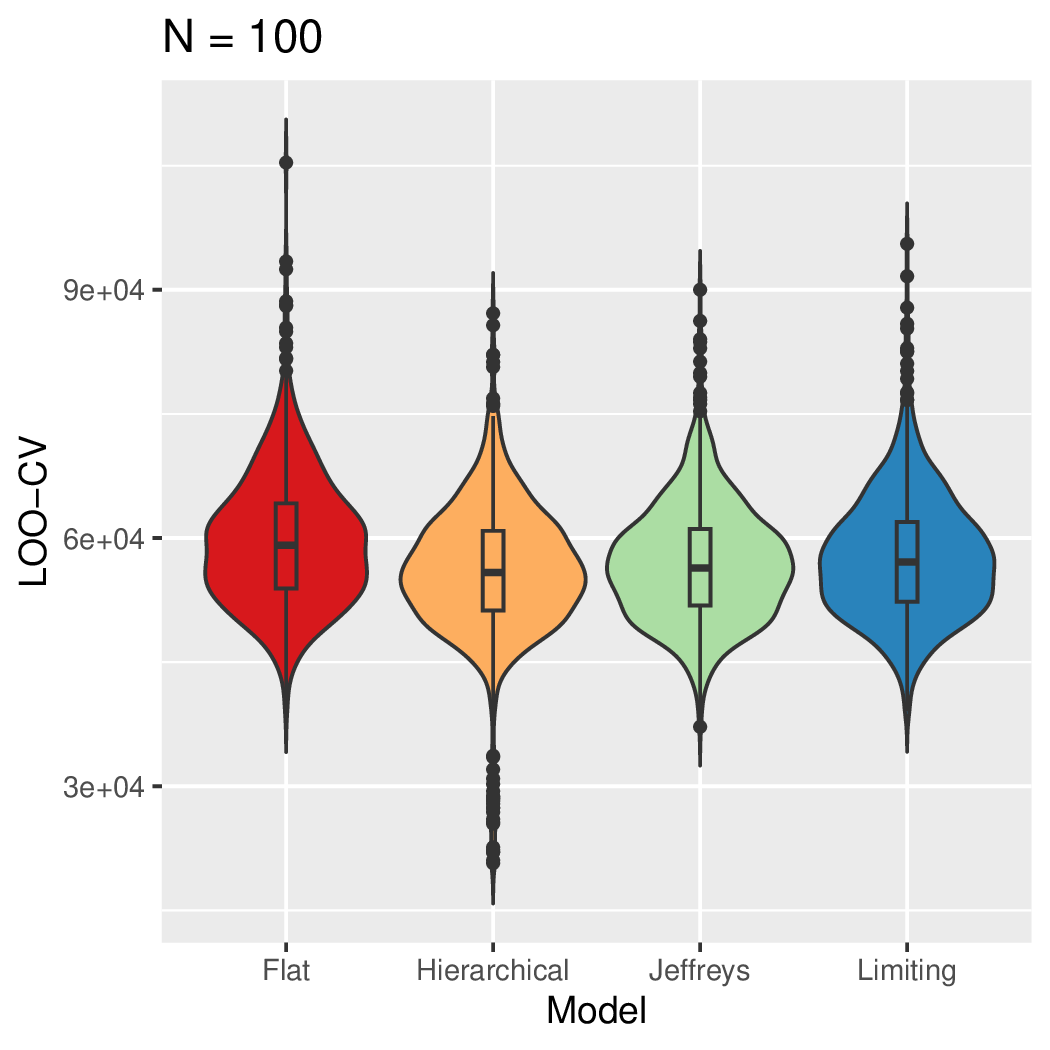}
	\includegraphics[scale = 0.35]{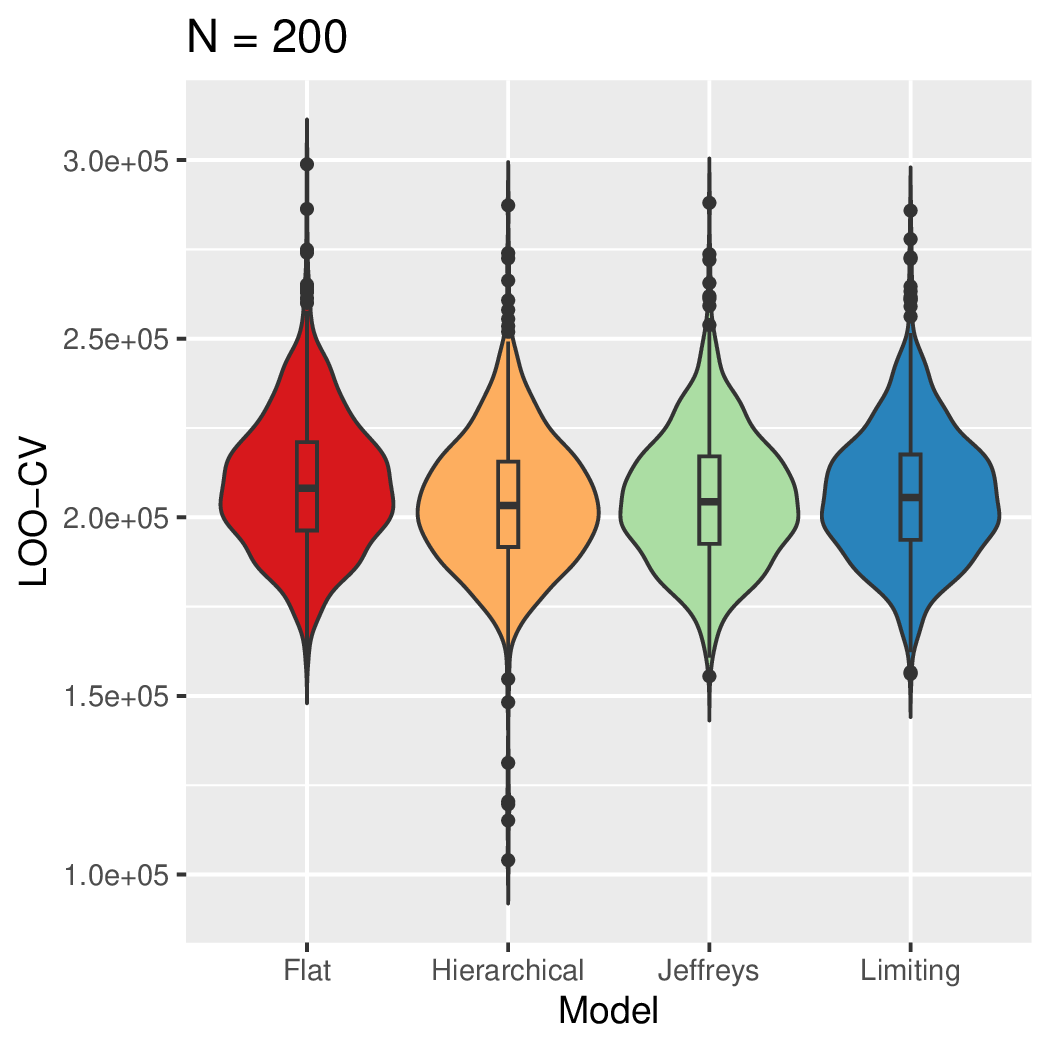}
	\includegraphics[scale = 0.35]{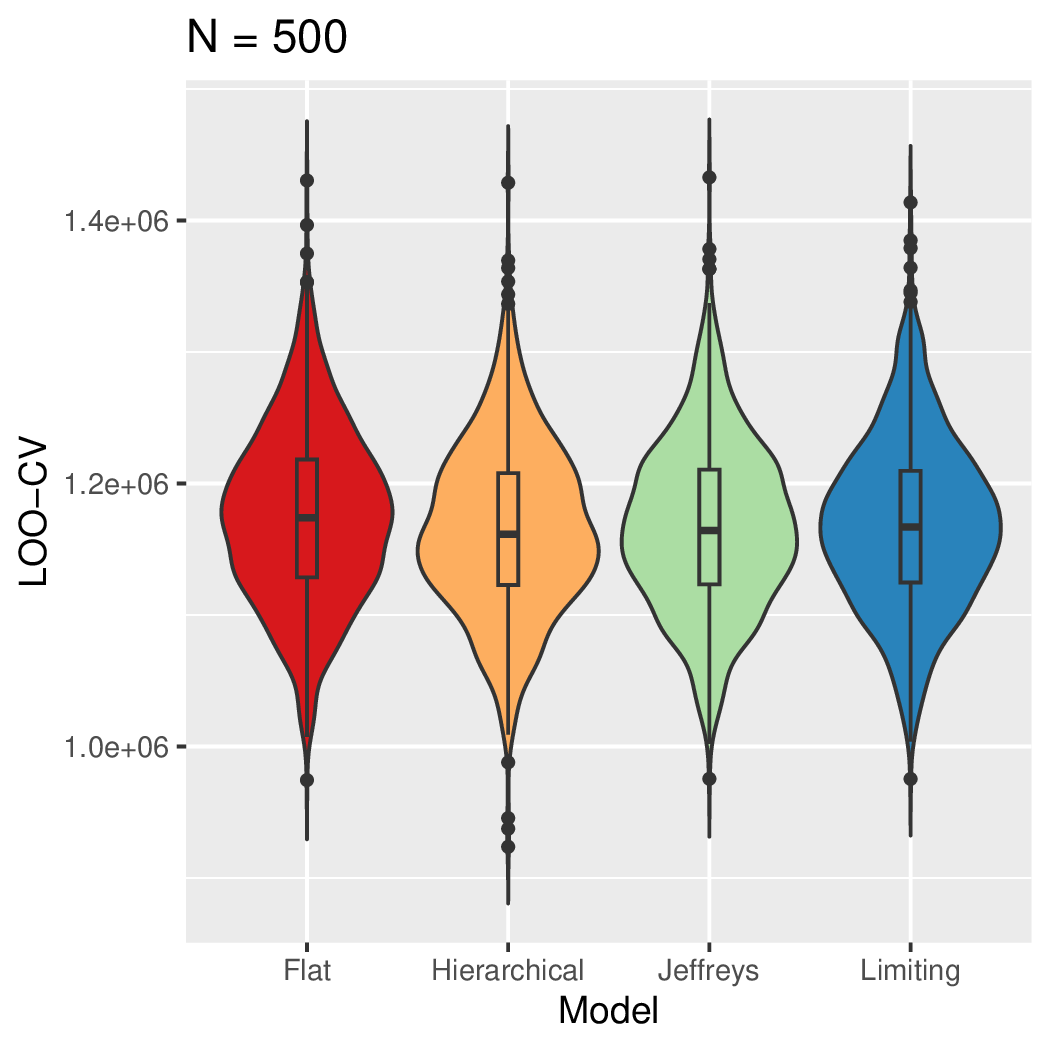}
	\caption{Leave-one-out Cross Validation (LOO-CV) for the underdispersed setting, by model and sample size.\label{f:uloos}}
\end{figure}

LOO-CV measures the predictive accuracy of the model and can be used to compare between posteriors resulting from different models, which we do here. Violin plots containing these values from all models are in Figures~\ref{f:ploos} to~\ref{f:uloos}. From these plots, we see that the model that uses the hierarchical reference prior consistently has the lowest LOO-CV across the three sample sizes for equidispersed and overdispersed data. As $N$ increases, the other models do attain LOO-CV scores closer to that of the hierarchical reference prior model, see Figures~\ref{f:ploos} and~\ref{f:oloos}. The differences are less pronounced in the underdispersed setting, however the hierarchal reference prior still tends to produce lower LOO-CVs across the three sample sizes.

%

\section{Data Illustrations}

To illustrate the use of the priors we consider, we present analyses of three different datasets. These illustrations include one equidispersed dataset on $\alpha$ particles emitted by polonium, one overdispersed dataset on horseshoe crab satellites, and one underdispersed dataset on the length of words in a Slovak poem. The sample sizes range from medium (117 and 173) to large (2,608). They come from a range of substantive fields illustrating the applicability of both the COM-Poisson model in general and the use of objective priors for an initial analysis.  Table~\ref{t:data} displays the posterior estimates (medians) and 95\% credible intervals (CrI) for both parameters estimated from each data set.  For each model, we draw 4,000 samples each from four separate chains, retaining the last 2,000 for a total of 8,000 retained posterior samples. As in the empirical study, we monitor convergence across the four chains using the potential scale reduction factor, $\hat{R}$ \citep{GelmanRubin1992}. Based on model performance of our empirical study in Section~\ref{s:sim}, we focus our discussion of the results on the hierarchical reference prior model. However, we also perform sensitivity analyses to the choice of prior for each illustration. All datasets and the scripts used to fit each model are available online at \url{https://github.com/markjmeyer/CMP}.

\begin{table}
	\centering
	\caption{Estimates (Est.; presented as posterior medians) and 95\% credible intervals (CrI) for all data illustrations. Each result is based on a mixture of four chains, each with 2000 retained posterior draws after a warmup of 2000 samples. \label{t:data}}
	\begin{tabular}{lcclcc}
	  \toprule
	Data & $n$ & Parameter & Model & Est. (CrI) & $\hat{R}$ \\ 
	  \midrule
	$\alpha$-Particles & 2608 & $\lambda$ & Flat & 4.1745 (3.7904, 4.5803) & 1.0038 \\ 
	  & & & Hierarchical & 4.1427 (3.7475, 4.5584) & 1.0027 \\ 
	  & & & Jeffreys & 4.1511 (3.7798, 4.5708) & 1.0028 \\ 
	  & & & Limiting & 4.1588 (3.7762, 4.5842) & 1.0060 \\ 
	  \cmidrule{3-6}
	& & $\nu$ & Flat & 1.0504 (0.9869, 1.1111) & 1.0040 \\ 
	 &  & & Hierarchical & 1.0451 (0.9796, 1.1078) & 1.0027 \\ 
	 &  & & Jeffreys & 1.0463 (0.9856, 1.1092) & 1.0024 \\ 
	 &  & & Limiting & 1.0478 (0.9832, 1.1130) & 1.0057 \\ 
	  \cmidrule{2-6}
	Crab Satellites & 173 & $\lambda$ & Flat & 0.7892 (0.7284, 0.9208) & 1.0078 \\ 
	  & & & Hierarchical & 0.7855 (0.7300, 0.8737) & 1.0040 \\ 
	  & & & Jeffreys & 0.7830 (0.7326, 0.8819) & 1.0006 \\ 
	  & & & Limiting & 0.7833 (0.7277, 0.8837) & 1.0035 \\ 
	  \cmidrule{3-6}
	& & $\nu$ & Flat & 0.0342 (0.0010, 0.1276) & 1.0072 \\ 
	  & & & Hierarchical & 0.0320 (0.0014, 0.0993) & 1.0056 \\ 
	  & & & Jeffreys & 0.0282 (0.0013, 0.1003) & 1.0011 \\ 
	  & & & Limiting & 0.0298 (0.0010, 0.1081) & 1.0037 \\ 
	  \cmidrule{2-6}
	Slovak Poem & 117 & $\lambda$ & Flat & 116.58 (34.528, 479.63) & 1.0012 \\ 
	  & & & Hierarchical & 63.520 (22.874, 214.53) & 1.0031 \\ 
	  & & & Jeffreys & 72.988 (24.765, 255.88) & 1.0029 \\ 
	  & & & Limiting & 77.928 (25.205, 294.87) & 1.0042 \\ 
	  \cmidrule{3-6}
	& & $\nu$ & Flat & 4.0038  (3.0263, 5.1345) & 1.0015 \\ 
	  & & & Hierarchical & 3.5125 (2.6788, 4.4938) & 1.0048 \\ 
	  & & & Jeffreys & 3.6293 (2.7451, 4.6431) & 1.0028  \\ 
	  & & & Limiting & 3.6748 (2.7603, 4.7475) & 1.0044 \\ 
	   \toprule
	\end{tabular}
\end{table}

\subsection{$\alpha$ Particles}


The largest dataset concerns the number of $\alpha$-particles emitted from a radioactive substance, in this case polonium, over 2,608 separate time intervals \citep{Rutherford1910}. In this experiment, the number of $\alpha$ particles emitted in 7.5 $s$ time intervals was counted using the scintillation method. The counts ranged from zero to 14 and the data is described in the table on page 701 of the original publication which sought to characterize the distribution of $\alpha$ particle emission \citep{Rutherford1910}. The estimates of the prior parameters for the hierarchical reference prior are $a = 0.593, b = 1.602, $ and $c = 0.232$. From our analysis in Table~\ref{t:data}, we note that the $\hat{R}$ values are all below 1.01 for both $\lambda$ and $\nu$ in each model, suggesting convergence. Focusing on the results from the hierarchical reference prior model, we observe that the estimate of $\lambda$ is a rate of $4.1427$ [95\% CrI $(3.7475, 4.5584)$] $\alpha$-particles emitted per 7.5 $s$. The estimate of $\nu$ is $1.0451$ [95\% CrI $(0.9796, 1.1078)$] which provides evidence of equidispersion in the data. Examining the estimates from the other models, there is a consensus on the emission rate being around 4.15. The CrIs for $\lambda$ are all fairly consistent with each other across models as well. The models also agree on the dispersion parameter being around 1.05. Moreover, the 95\% CrIs on $\nu$ all provide evidence of equidispersion in the sample.


\subsection{Crab Satellites}


A common example of overdispersion comes from a study of the mating patterns of horseshoe crabs on an island in the Gulf of Mexico \citep{Brockmann1996,Agresti2013}. Horseshoe crabs mate in pairs with the female crab arriving on shore with the male attached. Unattached males may group around a mating pair in clusters known as satellites and attempt to fertilize the eggs as well. The study examined different characteristics that contribute to the number of satellites near a female horseshoe crab's nest, including the female crab's color, spine condition, weight, and carapace width. In total, there are 173 female crabs available for study. Our analysis focuses simply on the count of satellites. The number of satellites per female crab ranges from zero to 15. We estimate the prior parameters for the hierarchal reference prior model to be $a = 0.2690, b = 1.6574,$ and $0.0007$. The $\hat{R}$ values in Table~\ref{t:data} suggest convergence for both model parameters across three models while the independent Jeffreys' prior model is slightly above 1.01. From the hierarchical reference prior model in Table~\ref{t:data}, the estimated value of the dispersion parameter $\nu$ is $0.0320$ [95\% CrI $(0.0014, 0.0993)$] which suggests the data is overdispersed. The estimated rate $\lambda$ is $0.7855$ [95\% CrI $(0.7300, 0.8737)$] satellites per nesting pair. Switching priors to one of our three options does not impact the results much. The estimate of the rate remains around 0.78 while the estimated dispersion is around 0.03. The CrIs are also similar and, in the case of $\nu$, provide evidence to suggest the data is overdispersed.

\subsection{Slovak Poem}

The distributions of word lengths based on different writing examples often result in underdispersion. One such example comes from the Slovak poem ``Ve\u{c}er po pr\'{a}ci'' by M. R\'{u}fus which contains 117 words of varying length, ranging from one to five \citep{Wimmer1994}. For the hierarchical reference prior model, the estimated prior parameters are $a = 1.273, b = 2.027,$ and $c = 0.1307$. The $\hat{R}$ values in Table~\ref{t:data} suggest convergence for both parameters across all models with all $\hat{R} < 1.01$. The hierarchical reference prior estimate of $\lambda$ is 62.520 [95\% CrI $(22.874, 214.53)$. The dispersion parameter estimate and CrI, 3.5125 [95\% CrI $(2.6788, 4.4938)$], provide evidence that the data is underdispersed. Comparing to the remaining models in Table~\ref{t:data}, we see consistency across models for the estimate of $\nu$, between 3.5 and 4 with somewhat similar CrIs. The estimated rate, however, exhibits a little more variability. While the hierarchical reference prior, independent Jeffreys' prior, and limiting prior provided similar results in the range of 63 to 77, the joint flat prior estimate is 116.58. The CrIs for these parameters are quite large and encompass each other's estimates providing some consensus to the range of plausible values.


\section{Discussion}

Objective priors are an important tool for modeling in the absence of prior knowledge, as a ``first step'' analysis, or as a sensitivity analysis. Complex distributions with multivariate parameter spaces, like the COM-Poisson, require extra care when developing objective priors. As has been well established in the literature, the multivariate Jeffreys' prior comes up short in models with two or more parameters when the information matrix is not diagonal. Thus, we describe and evaluate four candidate objective priors for the COM-Poisson. All models perform similarly in terms of rMSE and produce similar estimates in our data illustrations.

Both the joint flat prior and joint limiting models are improper priors that rely on the behavior of the observed data to ensure regularity, as we establish in Theorems~\ref{th:flat} and~\ref{th:limit}. Provided one has enough data, these model should not encounter issues. They are also somewhat more intuitive models: the joint flat prior may be an appealing representation of a total lack of knowledge regarding the parameters while the joint limiting prior's equivalence to the Poisson reference prior may be appealing for modelers who are unsure of the dispersion but operate from a ``working'' equidispersed assumption.

The independent Jeffreys' prior is based on the conditional prior approach \citep{Sun1998}. However, the resulting prior itself is complicated to evaluate and results in a posterior we can only examine empirically. While the hierarchical reference prior requires the optimization of a marginal likelihood, we have laid out a novel approach for eliciting this prior by leveraging the bound on the conjugate prior's normalizing constant and constrained nonlinear optimization. In our empirical study, it also performed the best in terms of minimizing the LOO-CV. The marginal likelihood is, however, difficult to draw samples from which lead us to implement a nonlinear constrained optimization routine. Given the speed of the routine and the advantage of the hierarchical reference prior model in simulation, it is our main recommendation for use as an objective prior. Since we derive the prior using the conjugate model and constrained optimization, the posterior is also guaranteed to be proper.

\bibliographystyle{abbrvnat} 
\bibliography{fullbib.bib}

@Manual{NLopt2008,
    author = {Steven G. Johnson},
    title = {The NLopt nonlinear-optimization package},
    year = {2008},
    url = {https://github.com/stevengj/nlopt},
}

@Manual{RStan2023,
    title = {{RStan}: the {R} interface to {Stan}},
    author = {{Stan Development Team}},
    year = {2023},
    note = {R package version 2.26.22},
    url = {https://mc-stan.org/}
  }

@Manual{nloptr2025,
    title = {nloptr: R Interface to NLopt},
    author = {J. Ypma and S. G. Johnson and A. Stamm and H. W. Borchers and D. Eddelbuettel and B. Ripley and K. Hornik and J. Chiquet and A. Adler and X. Dai and J. Ooms and T. Kalibera and M. Jagan},
    year = {2025},
    note = {R package version 2.2.1},
    url = {https://CRAN.R-project.org/package=nloptr},
  }

@incollection{Powell1994,
 author	= {M. J. D. Powell},
 editor 	= {S. Gomez and J.-P. Hennart},
 title 		= {A Direct Search Optimization Method That Models the Objective and Constraint Functions by Linear Interpolation},
 booktitle 	= {Advances in Optimization and Numerical Analysis},
 year 	= {1994},
 publisher	= {Springer Netherlands},
 address	= {Dordrecht, Netherlands},
 pages	={51--67},
 isbn		= {978-94-015-8330-5},
 doi		= {10.1007/978-94-015-8330-5_4}
}

@Article{Conway1962,
 author 	= {R. W. Conway and W. L. Maxwell},
 title 		= {A queuing model with state dependent service rates},
 journal 	= {Journal of Industrial Engineering},
 year 	= {1962},
 volume 	= {12},
 pages 	= {132--136}
}

@Techreport{Minka2003,
  author 		= {T. P. Minka and G. Shmueli and J. B. Kadane and S. Borle and P. Boatwright},
  title 		= {Computing with the {COM-Poisson} distribution},
  institution 	= {Department of Statistics, Carnegie Mellon University},
  number 		= {Technical Report 776},
  year 		= {2003}
}

@article{Lord2012,
 author 	= {D. Lord and S. D. Guikema},
 title 		= {The {Conway--Maxwell--Poisson} model for analyzing crash data},
 journal 	= {Applied Stochastic Models in Business and Industry},
 volume 	= {28},
 pages 	= {122--127},
 doi 		= {10.1002/asmb.937},
 year 	= {2012}
}

@Book{Agresti2013,
  author	= {A. Agresti},
  title 	= {Categorical Data Analysis},
  publisher	= {John Wiley \& Sons},
  address	= {Hoboken, NJ},
  edition	= {3$^{\text{rd}}$},
  year  	= {2013}
}

@article{Huang2021,
 author 	= {A. Huang and A. S. I. Kim},
 title 		= {Bayesian {Conway--Maxwell--Poisson} regression models for overdispersed and underdispersed counts},
 journal 	= {Communications in Statistics - Theory and Methods},
 volume 	= {50},
 pages 	= {3094--3105},
 year 	= {2021},
 doi 		= {10.1080/03610926.2019.1682162}
}

@book{Sellers2023, 
 author	= {K. F. Sellers}, 
 place 	= {Cambridge}, 
 series	= {Institute of Mathematical Statistics Monographs}, 
 title		= {The {Conway–Maxwell–Poisson} Distribution}, 
 doi 		= {10.1017/9781108646437}, 
 publisher	= {Cambridge University Press}, 
 address	= {Cambridge, United Kingdom},
 year		= {2023}, 
 collection = {Institute of Mathematical Statistics Monographs}
}

@article{Kadane2006,
 author 	= {J. B. Kadane and G. Shmueli and T. P. Minka and S. Borle and P. Boatwright},
 title 		= {{Conjugate analysis of the Conway-Maxwell-Poisson distribution}},
 volume 	= {1},
 journal 	= {Bayesian Analysis},
 pages 	= {363--374},
 year 	= {2006},
 doi 		= {10.1214/06-BA113}
}

@article{Chanialidis2018,
 author 	= {C. Chanialidis and L. Evers and T. Neocleous and A. Nobile},
 title 		= {Efficient {Bayesian inference for COM-Poisson} regression models},
 volume 	= {28},
 journal 	= {Statistics and Computing},
 pages 	= {595--608},
 year 	= {2018},
 doi 		= {10.1007/s11222-017-9750-x}
}

@article{Benson2021,
 author 	= {A. Benson and N. Friel},
 title 		= {Bayesian Inference, Model Selection and Likelihood Estimation using Fast Rejection Sampling: The {Conway-Maxwell-Poisson} Distribution},
 volume 	= {16},
 journal 	= {Bayesian Analysis},
 pages 	= {905--931},
 year 	= {2021},
 doi 		= {10.1214/20-BA1230}
}

@article{Jeffreys1946,
 author 	= {H. Jeffreys},
 title 		= {An invariant form for the prior probability in estimation problems},
 journal 	= {Proceedings of the Society A},
 volume 	= {186},
 pages 	= {453--461},
 year 	= {1946},
 doi 		= {10.1098/rspa.1946.0056}
}

@article{Bernardo1979,
 author 	= {J. M.Bernardo},
 title 		= {Reference Posterior Distributions for Bayesian Inference},
 journal 	= {Journal of the Royal Statistical Society: Series B},
 volume 	= {41},
 pages 	= {113--128},
 year 	= {1979},
 doi 		= {10.1111/j.2517-6161.1979.tb01066.x}
}

@article{Berger1989,
 author 	= {J. O. Berger and J. M. Bernardo},
 title 		= {Estimating a Product of Means: Bayesian Analysis with Reference Priors},
 journal 	= {Journal of the American Statistical Association},
 volume 	= {84},
 pages 	= {200--207},
 year 	= {1989},
 doi 		= {10.1080/01621459.1989.10478756}
}

@incollection{BergerBernardo1992BS4,
 author 	= {J. O. Berger and J. M. Bernardo},
 editor 	= {J. M. Bernardo and J. O. Berger and P. Dawid and A. F. M. Smith},
 isbn 	= {9780198522669},
 title 		= {On the Development of Reference Priors},
 booktitle 	= {Bayesian Statistics 4: Proceedings of the Fourth Valencia International Meeting, Dedicated to the memory of Morris H. DeGroot, 1931--1989},
 publisher 	= {Oxford University Press},
 address 	= {Oxford, United Kingdom},
 year 	= {1992},
 month 	= {08},
 doi 		= {10.1093/oso/9780198522669.003.0003}
}

@article{BergerBernardo1992,
 author 	= {J. O. Berger and J. M. Bernardo},
 title 		= {Ordered group reference priors with application to the multinomial problem},
 journal 	= {Biometrika},
 volume 	= {79},
 pages 	= {25--37},
 year 	= {1992},
 doi 		= {10.1093/biomet/79.1.25}
}

@Article{GelmanRubin1992,
 author	= {A. Gelman and D. B. Rubin},
 title		= {Inference from iterative simulation using multiple sequences},
 journal	= {Statistical Science},
 volume	= {7},
 pages	= {457--511},
 doi 		= {10.1214/ss/1177011136},
 year		= {1992}
}

@article{Gilks1992,
 author 	= {Gilks, W. R. and Wild, P.},
 title 		= {Adaptive Rejection Sampling for {Gibbs} Sampling},
 journal 	= {Journal of the Royal Statistical Society, Series C},
 volume 	= {41},
 pages 	= {337--348},
 year 	= {1992},
 doi 	= {10.2307/2347565}
}

@incollection{Neal1996,
 author	= {R. M. Neal},
 booktitle	= {Bayesian Learning for Neural Networks},
 title 		= {{Monte Carlo} Implementation},
 publisher 	= {New York: Springer},
 address	= {New York, United States},
 pages 	= {55--98},
 doi 		= {10.1007/978-1-4612-0745-0_3},
 year  	= {1996}
}

@article{Sun1998,
 author 	= {D. Sun and J. O. Berger},
 title 		= {Reference priors with partial information},
 journal 	= {Biometrika},
 volume 	= {85},
 pages 	= {55--71},
 year 	= {1998},
 doi 	= {10.1093/biomet/85.1.55}
}

@incollection{Bernardo2005,
 title 		= {Reference Analysis},
 editor 	= {D.K. Dey and C.R. Rao},
 series 	= {Handbook of Statistics},
 publisher 	= {Elsevier/North-Holland},
 volume 	= {25},
 pages 	= {17--90},
 year 	= {2005},
 booktitle 	= {Bayesian Thinking},
 address	= {Amsterdam, Netherlands},
 issn 		= {0169-7161},
 doi 		= {https://doi.org/10.1016/S0169-7161(05)25002-2},
 author 	= {J. M. Bernardo}
}

@article{Moller2006,
 author 	= {J. M{\o}ller and A. N. Pettitt and R. Reeves and K. K. Berthelsen},
 title 		= {An efficient {Markov chain Monte Carlo} method for distributions with intractable normalising constants},
 journal 	= {Biometrika},
 volume 	= {93},
 pages 	= {451--458},
 year 	= {2006},
 doi 		= {10.1093/biomet/93.2.451}
}

@article{Berger2009,
 author 	= {J. O. Berger and J. M. Bernardo and D. Sun},
 title 		= {The formal definition of reference priors},
 volume 	= {37},
 journal 	= {The Annals of Statistics},
 pages 	= {905--938},
 year 	= {2009},
 doi = {10.1214/07-AOS587}
}

@article{Ghosh2011,
 author 	= {M. Ghosh},
 title 		= {{Objective Priors: An Introduction for Frequentists}},
 volume 	= {26},
 journal 	= {Statistical Science},
 pages 	= {187--202},
 year 	= {2011},
 doi 		= {10.1214/10-STS338}
}

@article{Berger2012,
 author 	= {J. O. Berger and J. M. Bernardo and D. Sun},
 title 		= {Objective Priors for Discrete Parameter Spaces},
 journal 	= {Journal of the American Statistical Association},
 volume 	= {107},
 pages 	= {636--648},
 year 	= {2012},
 doi 		= {10.1080/01621459.2012.682538}
}

@article{Hoffman2014,
 author 	= {M. D. Hoffman and A. Gelman},
 title 		= {The {No-U-Turn Sampler}: Adaptively Setting Path Lengths in {Hamiltonian Monte Carlo}},
 journal 	= {Journal of Machine Learning Research},
 volume 	= {15},
 pages 	= {1593--1623},
 year 	= {2014}
}

@article{Berger2015,
 author 	= {J. O. Berger and J. M. Bernardo and D. Sun},
 title 		= {Overall objective priors},
 volume 	= {10},
 journal 	= {Bayesian Analysis},
 publisher 	= {International Society for Bayesian Analysis},
 pages 	= {189--221},
 year 	= {2015},
 doi 		= {10.1214/14-BA915}
}

@article{Rutherford1910,
 author 	= { E. Rutherford and  H. Geiger and H. Batemen},
 title 		= {The probability of variations in the distribution of $\alpha$ particles},
 journal 	= {The London, Edinburgh, and Dublin Philosophical Magazine and Journal of Science},
 volume 	= {20},
 pages 	= {698--701},
 year  	= {1910},
 doi 		= {10.1080/14786441008636955}
}

@article{Wimmer1994,
 author 	= { G. Wimmer  and  R. K\"ohler and R. Grotjahn  and  G. Altmann },
 title 		= {Towards a theory of word length distribution},
 journal 	= {Journal of Quantitative Linguistics},
 volume 	= {1},
 pages 	= {98-106},
 year  	= {1994},
 doi 		= {10.1080/09296179408590003}
}

@Article{Brockmann1996,
 author 	= {H. J. Brockmann},
 title 		= {Satellite Male Groups in Horseshoe Crabs, \emph{Limulus polyphemus}},
 journal 	= {Ethology},
 year 	= {1996},
 volume 	= {102},
 pages 	= {1--21},
 doi 		= {10.1111/j.1439-0310.1996.tb01099.x}
}

\end{document}